\documentclass[aps,pre,floatfix,twocolumn,superscriptaddress]{revtex4-2}
\usepackage[utf8]{inputenc}
\usepackage[english]{babel}
\usepackage{amsmath}
\usepackage{amsfonts}
\usepackage{amssymb}
\usepackage{graphicx}
\usepackage{color}
\usepackage{natbib}
\usepackage[colorlinks=true, citecolor=blue, linkcolor=blue, urlcolor=blue]{hyperref}
\usepackage{booktabs}
\usepackage{enumitem}
\usepackage{subfigure}
\usepackage{multirow}
\usepackage{braket}
\usepackage{dcolumn}
\usepackage{bm}
\usepackage[parse-numbers=true]{siunitx}
\usepackage{textgreek}
\usepackage{wasysym}
\usepackage{float} 
\usepackage{xcolor}
\usepackage{gensymb} 
\usepackage{comment}
\usepackage{pgfplots}
\usepackage{braket}
\usepackage[percent]{overpic}
\usepackage{epstopdf}

\begin{document}

\title{Physical properties of new delafossite triangular-lattice compounds TlErSe$_2$ and TlTmSe$_2$}

\author{Bastian Rubrecht}
    \affiliation{Institute for Solid State Research, Leibniz IFW Dresden, Helmholtzstra\ss{}e 20, 01069 Dresden, Germany}
    
\author{Ellen H\"au{\ss}ler}
    \affiliation{Faculty of Chemistry and Food Chemistry, Technische Universit\"at Dresden, 01062 Dresden, Germany}
    
\author{Mirtha Pillaca}
    \affiliation{Faculty of Chemistry and Food Chemistry, Technische Universit\"at Dresden, 01062 Dresden, Germany}
    
\author{Pritam Bhattacharyya}   
    \affiliation{Institute for Theoretical Solid State Physics, Leibniz IFW Dresden, Helmholtzstra\ss{}e 20, 01069 Dresden, Germany}
    
\author{Liviu Hozoi}   
    \affiliation{Institute for Theoretical Solid State Physics, Leibniz IFW Dresden, Helmholtzstra\ss{}e 20, 01069 Dresden, Germany}

\author{Artem Nosenko}   
    \affiliation{Institute for Solid State Research, Leibniz IFW Dresden, Helmholtzstra\ss{}e 20, 01069 Dresden, Germany}
    \affiliation{Kyiv Academic University, 03142 Akademika Vernads'koho Blvd., 36, Kyiv, Ukraine}
    
\author{Dmitri V. Efremov}   
    \affiliation{Institute for Solid State Research, Leibniz IFW Dresden, Helmholtzstra\ss{}e 20, 01069 Dresden, Germany}
    
\author{Bernd B\"uchner}
    \affiliation{Institute for Solid State Research, Leibniz IFW Dresden, Helmholtzstra\ss{}e 20, 01069 Dresden, Germany}
    \affiliation{Institute of Solid State and Material Physics and W\"urzburg-Dresden Cluster of Excellence ct.qmat, Technische Universit\"at Dresden, 01062 Dresden, Germany}

\author{Anja U.B. Wolter}
    \affiliation{Institute for Solid State Research, Leibniz IFW Dresden, Helmholtzstra\ss{}e 20, 01069 Dresden, Germany}

\author{Thomas Doert}
    \affiliation{Faculty of Chemistry and Food Chemistry, Technische Universit\"at Dresden, 01062 Dresden, Germany}

\date{\today} 

\begin{abstract}

Delafossite compounds containing rare-earth ions have been proven to be an ideal platform to investigate frustrated magnetic ground states. Here, we discuss two triangular-lattice antiferromagnets, TlErSe$_2$ and TlTmSe$_2$, as potential candidates for hosting exotic quantum states. Powder X-ray diffraction data analysis of the black-color polycrystalline Tl$RE$Se$_2$ ($RE$: Er and Tm) samples confirms the phase purity. Both materials crystallize in the trigonal $\alpha$-NaFeO$_2$ structure ($R\Bar{3}m$) with lattice parameters $a$ = 4.1070(4)\,\AA{} and $c$ = 23.1472(1)\,\AA{} for the erbium compound and $a$ = 4.0916(1)\,\AA{} and $c$ = 23.1483(2)\,\AA{} for the thulium compound. Magnetic susceptibility measurements show an effective moment of $\mu_{\mbox{eff}} = 9.6(2)\,\mu_B$/f.u. ($7.5(1)\,\mu_B$/f.u.) for TlErSe$_2$ (TlTmSe$_2$) for temperatures above 200\,K. While $^3$He specific-heat measurements reveal long-range magnetic order below $T_N = 0.42~$K for TlErSe$_2$, no sign of long-range magnetic order was observed for TlTmSe$_2$. Based on our results, we map out the T-H phase diagram for polycrystalline TlErSe$_2$ and discuss the striking difference in the magnetic behavior of TlTmSe$_2$ based on our ab initio quantum chemical calculations.
\end{abstract}

\maketitle

\section{Introduction}

Frustration lies at the heart of a variety of exotic magnetic ground states and responses in solid state magnetism. It can be of pure geometric nature as for the Heisenberg antiferromagnetic triangular lattice or can arise from competing anisotropic interactions as in Kitaev's honeycomb model. In some systems it may actually imply both geometric frustration and peculiar anisotropic couplings, e.\,g., on triangular lattices of edge-sharing MX$_6$ octahedra, where M stands for a $d$- or $f$-electron ion and X for chalcogen or halide species. Illustrative examples of such edge-sharing networks of MX$_6$ octahedra are triangular Kitaev-Heisenberg like CoI$_2$ \cite{Co_c.kim}, NaRuO$_2$ \cite{Ru_Pritam}, and CsCeSe$_2$ \cite{Ce_Xie} or Ising antiferromagnets with `intrinsic' transverse fields like TmMgGaO$_4$ \cite{Shen2019intertwined, Dun2021neutron,Hu2020Berezinskii,liu2020intrinsic} and KTmSe$_2$ \cite{expt_KTm}. The former provide the playground for studying quantum spin liquid (QSL) physics \cite{Ru_Pritam}, the latter are of interest in relation to Berezinskii-Kosterlitz-Thouless (BKT) phases \cite{PhysRevB.103.104416,PhysRevB.68.104409}.

Rare-earth delafossites with the chemical formula AMX$_2$ emerge in this context as a versatile material platform that allows to investigate both QSL and BKT states. Though the role of Kitaev exchange is not yet fully clarified in some cases, potential QSL ground states were pointed out in both Yb$^{3+}$ 4$f^{13}$ \cite{Baenitz2018,Ranjith2019NaYbO2,Sichelschmidt_2019,Ranjith2019NaYbSe2,Ding2019NaYbO2,bordelon2019field,Bordelon2020NaYbO2,Zhang2021NaYbSe2,Dai2021NaYbSe2,liu2018rare,lkqb-3fc7} and Ce$^{3+}$ 4$f^{1}$ \cite{Ce_ortiz} delafossites while the structurally related AMX$_2$ 4$f^{12}$ compounds represent promising potential BKT systems, since ambiguities arising from Mg-Ga intermixing and associated disorder as encountered in the TmMgGaO$_4$ material \cite{Shen2019intertwined, Dun2021neutron,Hu2020Berezinskii,liu2020intrinsic} are circumvented in the delafossite architecture.

Here we report the synthesis and investigation of both odd- and even-electron new compounds in the 4$f^n$ delafossite series, displaying the same ligand and A-site matrix --- TlErSe$_2$ and TlTmSe$_2$. The former features $\tilde{S}\!=\!1/2$ magnetic moments related to a Kramers-doublet ground state at each 4$f^{11}$ site and is found to order antiferromagnetically below 420 mK. No sharp phase transition is observed in the latter material; its low-energy multiplet structure, characterized by two nearly degenerate singlet states at each 4$f^{12}$ center, is discussed in relation to the transverse field Ising model \cite{Hu2020Berezinskii,PhysRevB.68.104409,liu2020intrinsic}. Our results add new reference points within the landscape of late rare-earth triangular-lattice magnetic systems. Studies on A=Tl delafossites, in particular, are scarce, likely due to the toxicity of thallium --- besides early works focusing on structural aspects or which were limited to temperature $T\geq 4.2~$K \cite{kabre1974crystallographic,duczmal1994synthesis,duczmal1995magnetic}, only the magnetism of the 4$f^{13}$ Kramers-ion systems TlYbS$_2$ \cite{Ferreira2020} and TlYbSe$_2$ \cite{lkqb-3fc7} has been addressed so far. Our work provides valuable insights into this particular direction.

\section{Experimental and computational methods}
\label{sec:methods}    
\textit{Synthesis. }
    Due to the air and moisture sensitivity of the starting materials and products, all the experimental work was carried out in an Ar-filled glove box with O$_2$ and H$_2$O levels below 0.1 ppm.
    Polycrystalline TlErSe$_2$ and TlTmSe$_2$ samples were synthesized in two steps via the solid state method by using high purity elements of Tl (Alfa Aesar, 99.9\%), Erbium (Mateck, 99.5 \%), Thulium (Mateck, 99.99 \%), and Se (Chempur, 99.999\%, reduced in H$_2$ stream at $400$\,$^\circ$C) as raw materials.   
    
    Firstly, the Tl$_2$Se precursor was synthesized from a stoichiometric mixture 2:1 of Tl and Se in an evacuated ($\sim10^{-3}$\,mbar) sealed quartz ampule. The ampule was placed in a muffle furnace (Nabertherm),  heated up to $500$\,$^\circ$C with a ramp of $100$\,K/h, kept at this temperature for $24$\,h, and finally cooled down to room temperature within $5$ hours. 
    
    In the second step, polycrystalline TlErSe$_2$ and TlTmSe$_2$ were synthesized out of a stoichiometric mixture of Tl$_2$Se, Se, and Er or Tm, respectively. The three components were weighted in a molar ratio of 1:2:3, mixed in an agate mortar, and placed in a glassy-carbon crucible inside a quartz ampule of $16$\,mm diameter. The evacuated ampule was heated to $500-540$\,$^\circ$C with a heating rate of $100$\,K/h and dwelled for $24$ hours. Afterward, the ampule containing TlTmSe$_2$ was cooled down with 10\,K/h, and the ampule containing TlErSe$_2$ was water quenched. Both obtained samples were pulverized for further structural and physical properties characterization.
    
    In addition, a non-magnetic reference material, TlLuSe$_2$, was synthesized in polycrystalline form following a similar procedure.  

\vspace{0.2cm}
\textit{Powder X-Ray diffraction}. 
    The phase purity of the prepared samples was checked by powder X-ray diffraction (PXRD) with an Empyrean diffractometer (PANalytical, Cu-K${\alpha_1}$ radiation) equipped with a Ge(111) monochromator. The data were collected in a Bragg-Brentano setup in the $2\theta$-range of 5 to 90$^{\circ}$. For the accurate lattice parameters measurement, the samples were mixed with a small amount of Si powder (NIST SRM $640c$, $a$ = 5.43119(1)\AA) as an internal standard. Lattice parameters of the samples were determined by Rietveld refinement using the TOPAS Academic software \cite{Coelho2018,Coelho2020}.

\vspace{0.2cm}
\textit{Magnetization and heat capacity}. 
    Magnetization and specific heat measurements were performed on pressed pellets of the polycrystalline sample. DC magnetometry was performed using a Superconducting QUantum Interference Device (SQUID) magnetometer from Quantum Design (MPMS 3) for fields up to $7~$T and a Physical Property Measurement System (PPMS) for magnetic fields up to $14~$T, both equipped with a Vibrating Sample Magnetometer (VSM) option. 
    
    The specific heat was measured using the relaxation method with a PPMS equipped with a $^3$He refrigerator option. The background signal of the sample holder was measured separately and then subtracted. The lattice contribution to the specific heat was approximated by the Lindeman-scaled isostructural non-magnetic analogue TlLuSe$_2$. 

\vspace{0.2cm}
\textit{Band structure DFT computations}. 
    First-principles calculations were conducted to investigate the electronic structures of TlErSe$_2$ and TlTmSe$_2$. 
    The ab initio simulations were performed using the projector augmented-wave (PAW) method  \cite{Kresse1999}, as implemented in the Vienna Ab initio Simulation Package (VASP) \cite{Kresse1996,Kresse1993}. An energy cutoff of $700~$eV was applied to the plane-wave basis set. The Brillouin zone was sampled using a $\Gamma$-centered  28x28x28 k-point mesh. Spin–orbit coupling (SOC) was included self-consistently in all calculations. The $ f $-electrons of Er and Tm were treated as core electrons. For both compounds the trigonal crystal structure (sg 166) together with the experimentally obtain lattice parameters were used. The crystal structure parameters are $a = 4.0916(1) $\AA\, and $c = 23.1483(2) $\AA\, for TlTmSe$_2$ and 
    $a = 4.1070(4) $\AA\, and $c = 23.1472(1) $\AA\, for TlErSe$_2$. All calculations assumed non-magnetic configurations. For post-processing we used the pymatgen python library \cite{Pymatgen}. 

\vspace{0.2cm}
\textit{Quantum chemical calculations}. 
{\it ab initio} quantum chemical calculations for the multiplet structures of both compounds, TlErSe$_2$ and TlTmSe$_2$, were carried out using the {\sc orca} \cite{ORCA_5,ORCA_5_1} and {\sc molpro} \cite{Molpro} programs, respectively. For this purpose, clusters consisting of one central ErSe$_6$/TmSe$_6$ octahedron, six adjacent Er/Tm sites, and 12 Tl nearest-neighbor cations were considered. The crystalline environments were modeled as large arrays of point charges which reproduce the crystalline Madelung field within the cluster volume; we employed the {\sc ewald} program \cite{Klintenberg_et_al,Derenzo_et_al} to generate the point-charge embeddings. The performance of such an embedded-cluster quantum chemical approach for the case of 4$f$-electron compounds is discussed in e.\,g.~\cite{NaCeO2,NaYbSe2}. 

The quantum chemical investigation was initiated as complete active space self-consistent field (CASSCF) computations \cite{olsen_bible} with active orbital spaces containing the seven 4$f$ orbitals of the central Er/Tm ion. Post-CASSCF calculations were performed for the TlTmSe$_2$ system at the level of multireference configuration-interaction (MRCI) with single and double excitations \cite{olsen_bible,MRCI_Molpro} out of the Tm 4$f$ and Se 4$p$ orbitals of the central TmSe$_6$ octahedron. The spin-orbit (SO) interaction matrix was computed for the lowest 11 spin triplet (out of a total number of 21) and nine spin singlet (out of a total number of 28) 4$f^{12}$ states. Other higher-lying triplet and singlet 4$f^{12}$ states have relative energies that are respectively larger by $\approx$1 and $\approx$2.4 eV, in the calculation without SOC; excluding those from the SO treatment does not significantly affect the low-energy 4$f^{12}$ relativistic spectrum in TlTmSe$_2$, as illustrated at CASSCF level in the Appendix (Tab. \ref{tab:excited_TlTm_CAS}).

With three holes in the 4$f$ subshell, MRCI is computationally too expensive for the TlErSe$_2$ compound. Post-CASSCF calculations were therefore performed for TlErSe$_2$ at the level of $N$-electron valence second-order perturbation theory (NEVPT2) \cite{NEVPT2}. All possible states (35 quartets and 112 doublets) derived from the 4$f^{11}$ valence configuration were included in the SO treatment either at CASSCF or NEVPT2 level.

The basis sets used in the computational study are described in the Appendix.

\section{Results}
\begin{figure}
                \includegraphics[width=\columnwidth]{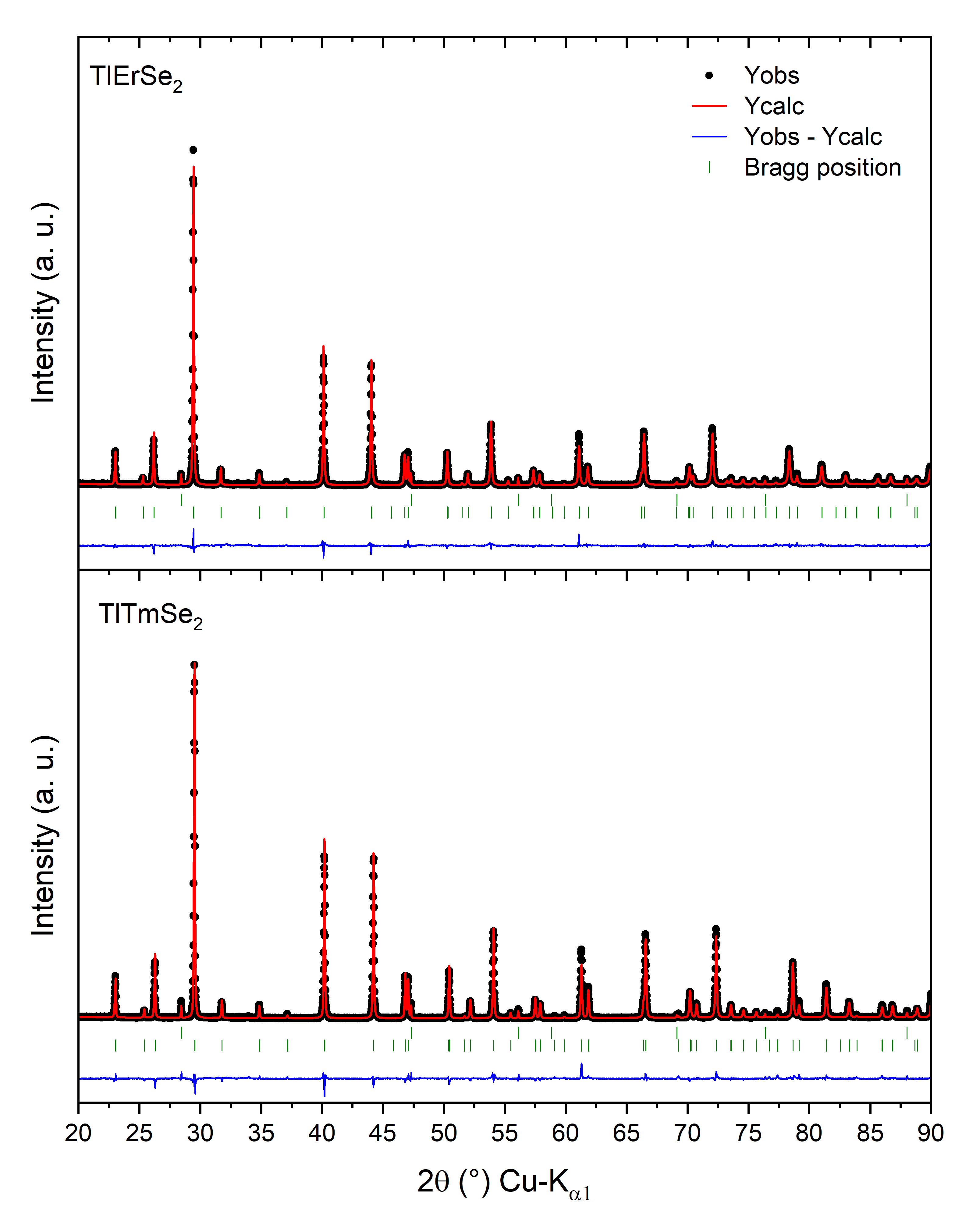}\\
                \caption{PXRD pattern with Rietveld refinement analysis for TlErSe$_2$ (top) and TlTmSe$_2$ (bottom) sample. Solid black circles are observed intensities, red lines are the calculated intensities, and blue lines are the difference between the observed and the calculated intensities. Short vertical lines show the position of Bragg reflections for TlErSe$_2$ and TlTmSe$_2$, respectively as well as for Si (internal standard).}
                \label{fig:TRES-xray}
        \end{figure}

        The powder X-ray diffraction pattern based Rietveld analysis, shown in Fig. \ref{fig:TRES-xray}, confirms the single-phase formation of the polycrystalline TlErSe$_2$ and TlTmSe$_2$ compounds with the rhombohedral $\alpha$-NaFeO$_2$ structure and space group $R\Bar{3}m$ (No. 166). Tl$^{1+}$, RE$^{3+}$, and Se$^{2-}$ atoms occupy the Wyckoff position (and atomic coordinates) 3$b$ (0, 0, 0.5), 3$a$ (0, 0, 0) and 6$c$ (0, 0, $z$), respectively. Therefore, during the refinement analysis, only the $z$ coordinate of the Se atom was refined. Lattice parameters as well as the reliability factors of both compounds obtained are shown in Table \ref{tab:struc-parameters}. They are in good agreement with the values reported by Duczmal \textit{et al.} \cite{duczmal1995magnetic,duczmal2003struktura}, but quite different from the ones reported by Kabre \textit{et al.} \cite{kabre1974crystallographic}.
        This difference may be related to the fact that in Ref. \cite{kabre1974crystallographic} the lattice parameters determination was done over $12$ diffraction peaks only. The lattice parameters of the isostructural and non-magnetic reference TlLuSe$_2$ are $a$ = 4.0656(7)\AA, $c$ = 23.1769(4)\,\AA{}, and $V$ = 331.77(1)\,\AA$^3$.
        The isotropic thermal displacement parameters B$_{eq}$ were kept fixed at 1.0 \AA$^2$ and the site occupancies were also fixed to 1.0 in the refinements and structural parameters of the Rietveld fits are summarized in Table \ref{tab:struc-parameters}.
        Furthermore, energy dispersive X-ray (EDX) analysis and inductively coupled plasma optical emission spectroscopy (ICP-OES) measurements in the powder sample of TlErSe$_2$ and TlTmSe$_2$ confirm the homogeneity and stoichiometric composition of the delafossite compounds. 

\begin{figure}
                \centering
                \includegraphics[width=\columnwidth]{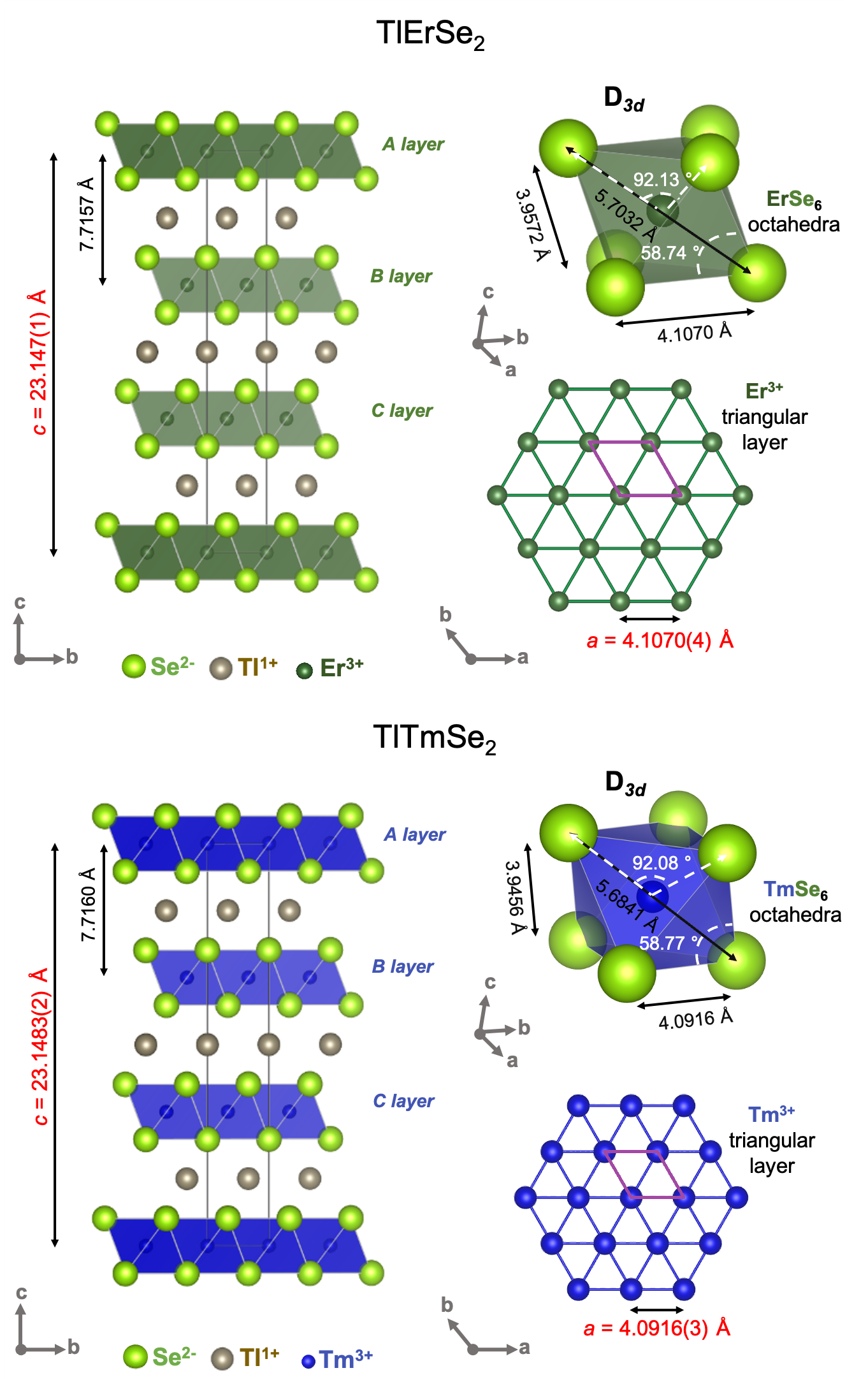}
                \caption{Refined crystal structures of  TlErSe$_2$ and TlTmSe$_2$ with space group $R\Bar{3}m$.  The ErSe$_6$ and TmSe$_6$ distorted-octahedra with $D_{3d}$ structure and the diagram of a single perfect triangular layer of  Er$^{3+}$ and Tm$^{3+}$ ions (viewed along the $c$-axis) are also shown. The unit cells are indicated with solid purple lines. Tl$^{1+}$, Er$^{3+}$/Tm$^{3+}$, and Se$^{2-}$ atoms are shown in brown, green/blue, and light green,  respectively. Illustrations were generated in VESTA \cite{Momma2011vesta}.}
                \label{fig:structure}
        \end{figure}

        The refined crystal structures of TlErSe$_2$ and TlTmSe$_2$ are shown in Fig. \ref{fig:structure}, in which the stacked layered ErSe$_6$ and TmSe$_6$ octahedra are arranged in an \textit{ABC} sequence and separated by $7.718~$\AA{} and $7.7116~$\AA{}, respectively. The bond lengths and angles of the slightly distorted ErSe$_6$ and TmSe$_6$ octahedra together with a single layer of Er/Tm atoms ordered in a perfect triangular lattices are also indicated.

        \begin{table*}
            \caption{Rietveld results of structural parameters, atomic coordinates ($x$, $y$, $z$), and equivalent isotropic displacement parameters (B$_{eq}$). Residual factors are $R_{p}$(\%) = 7.31, $R_{wp}$(\%) = 9.71, $R_{exp}$(\%) = 4.82, and the goodness-of-fit is $GoF$ = 2.02 for TlTmSe$_2$ and $R_{p}$(\%) = 5.40, $R_{wp}$(\%) = 7.10, $R_{exp}$(\%) = 4.89, and $GoF$ = 1.45 for TlErSe$_2$..}
            \renewcommand{\arraystretch}{1.3}
            \begin{ruledtabular}
            \begin{tabular}{cccccccc}
                Lattice parameters                &Atom  & Wyckoff position&$x$ &$y$ & $z$     &Occupancy & B$_{eq}$(\AA$^2$)\\ 
                \hline
                $a$ = 4.1070(4)\,\AA       &Er    &3$a$             & 0  & 0  &   0     & 1.0      & 1.0 \\
                $c$ = 23.147(1)\,\AA{}   &Tl    &3$b$             & 0  & 0  &  1/2    & 1.0      & 1.0 \\
                $V$ = 338.13(7)\,\AA$^3$ &Se    &6$c$             & 0  & 0  &0.267(1)& 1.0      & 1.0 \\
                \hline
                $a$ = 4.0916(1)\,\AA       &Tm    &3$a$             & 0  & 0  &   0     & 1.0      & 1.0 \\
                $c$ = 23.1483(2)\,\AA{}   &Tl    &3$b$             & 0  & 0  &  1/2    & 1.0      & 1.0 \\
                $V$ = 335.602(6)\,\AA$^3$ &Se    &6$c$             & 0  & 0  &0.2657(5)& 1.0      & 1.0 \\
            \end{tabular}
            \end{ruledtabular}
            \label{tab:struc-parameters}
        \end{table*}
    
        \begin{figure}
            \begin{overpic}[width=\linewidth]{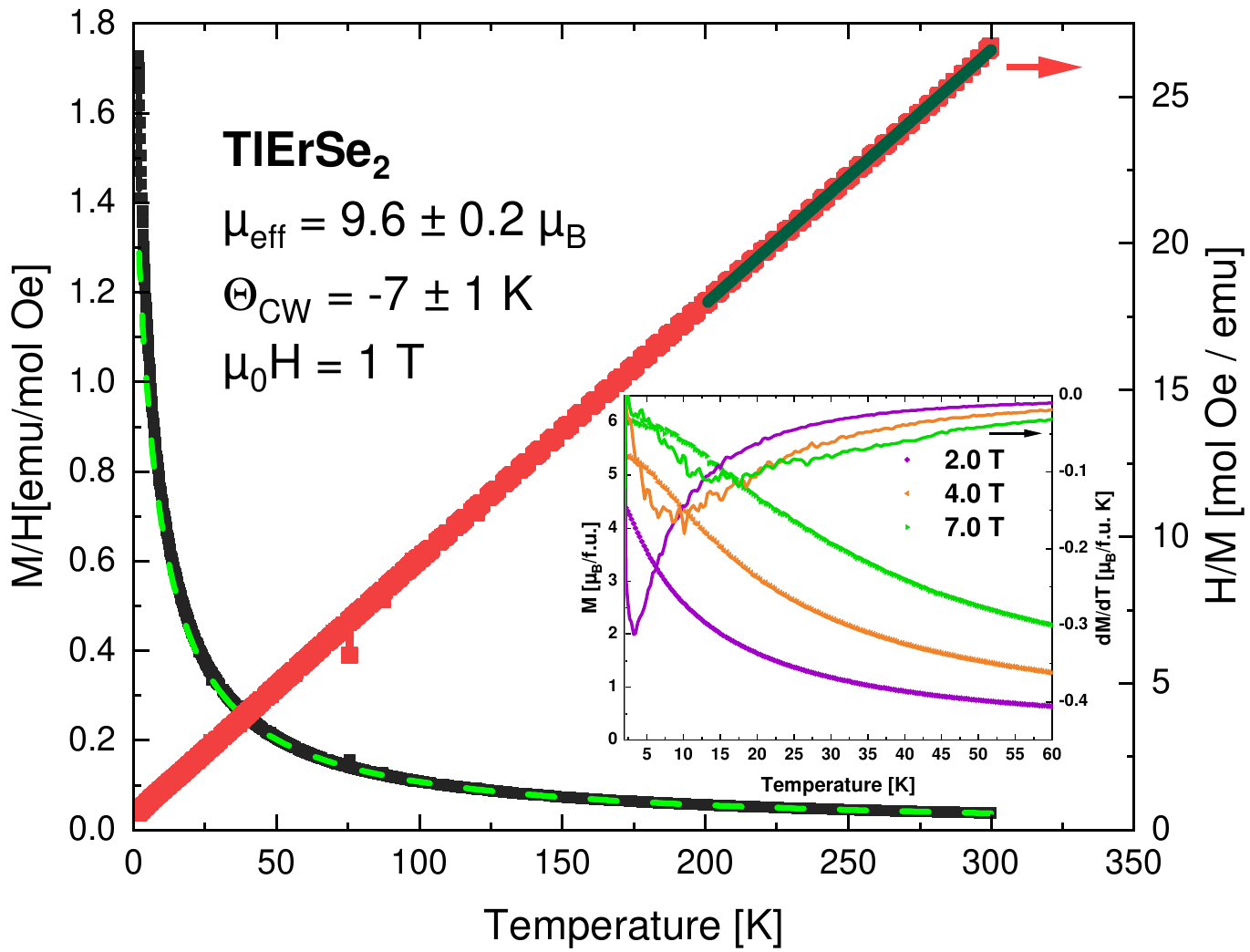}
                \put(1,72){\textbf{a)}}
            \end{overpic}
            \begin{overpic}[width=\linewidth]{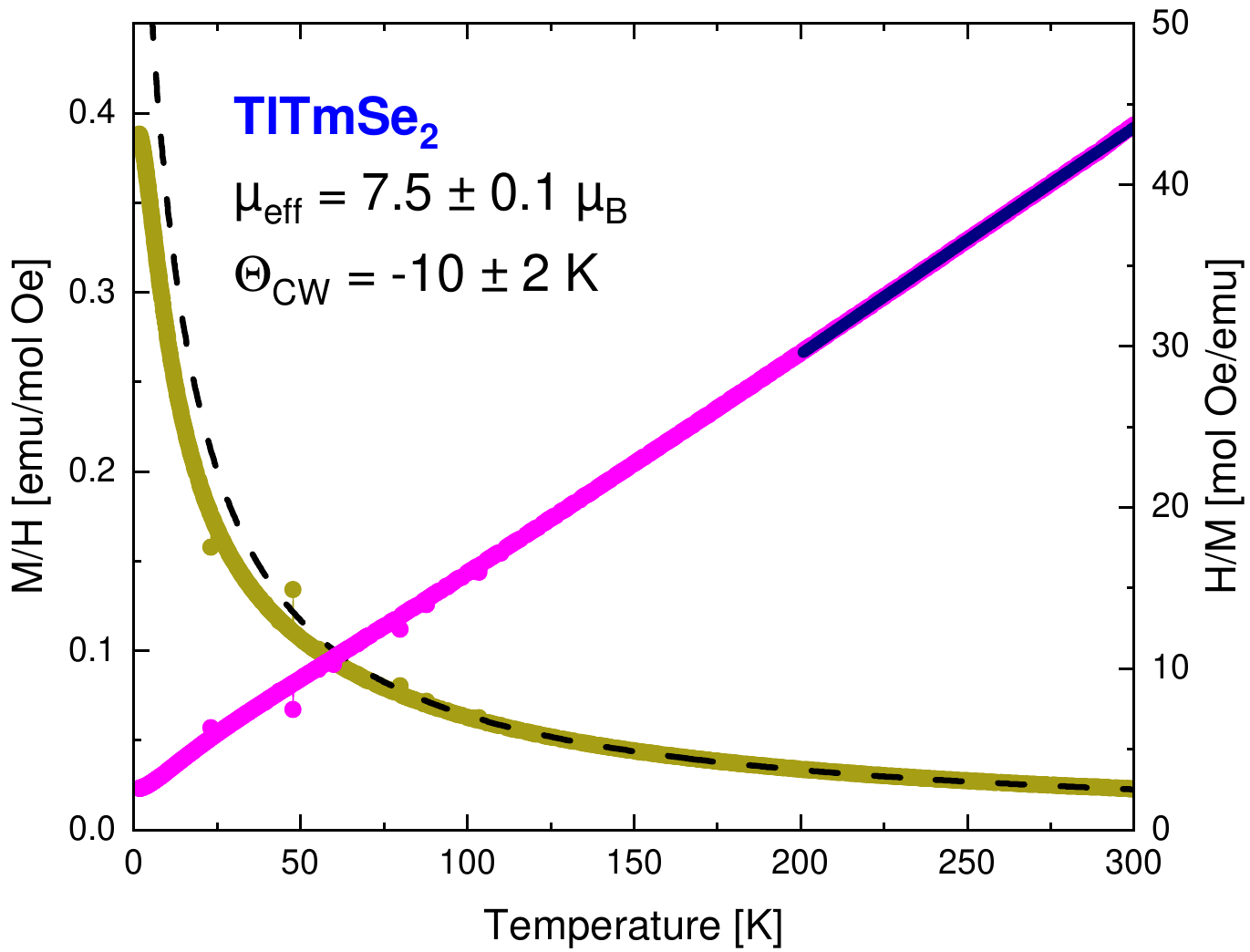}
                \put(1,72){\textbf{b)}}
            \end{overpic}
            \caption{a) Temperature dependence of the magnetization of TlErSe$_2$ divided by the applied field ($M/H$) and its inverse. The insert shows the temperature dependence of the magnetization and its first derivative of TlErSe$_2$ for applied fields of $2$, $4$, and $7$\,T. b) $M/H$ and its inverse of TlTmSe$_2$ as a function of temperature measured at $1$\,T. The solid lines correspond to a Curie-Weiss fit for temperatures above $200$\,K. 
            }
            \label{fig:TRES_MxT}
        \end{figure}

        \begin{figure}
            \centering
            \begin{overpic}[width=\linewidth]{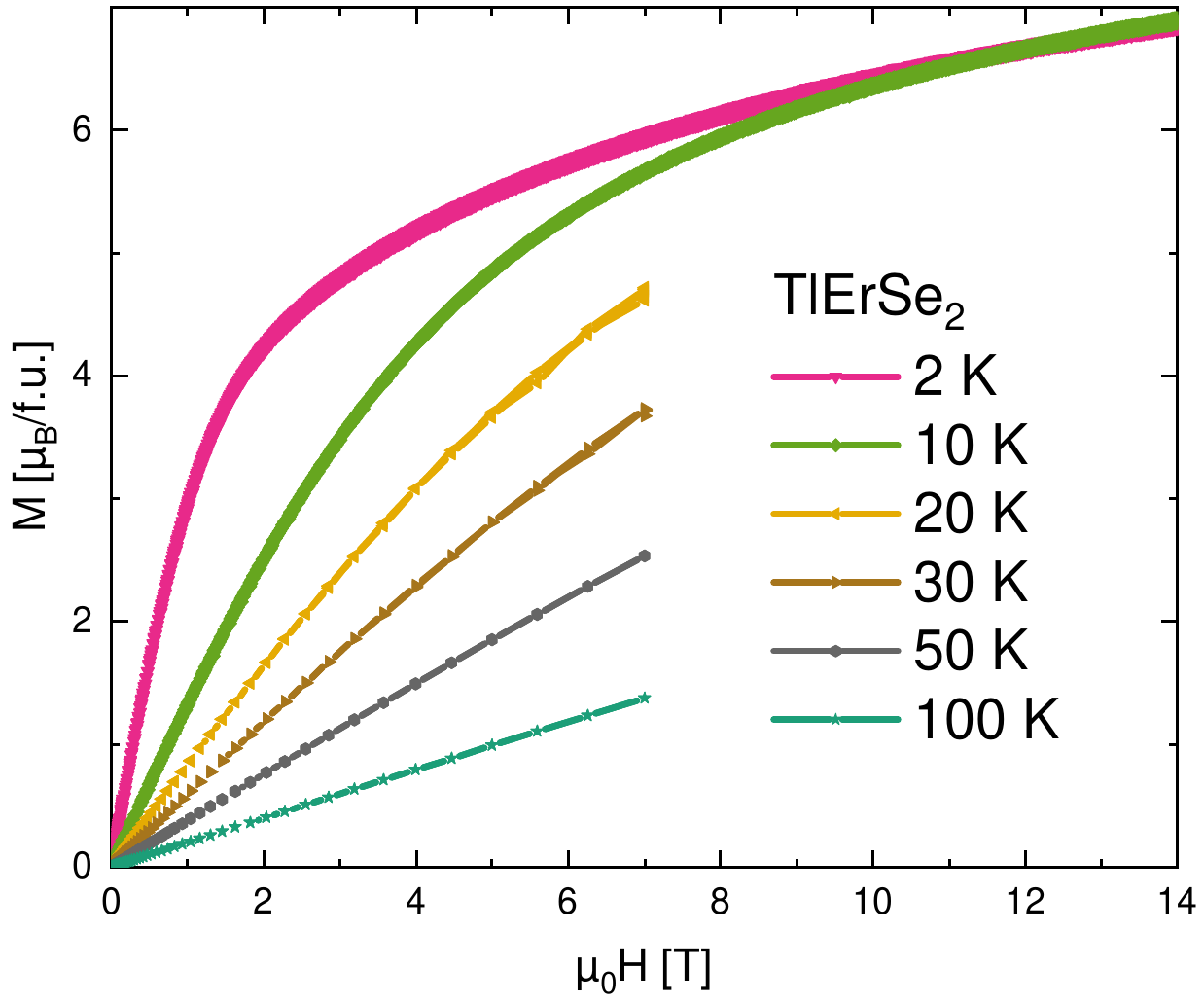}
                \put(1,80){\textbf{a)}}
            \end{overpic}
            \begin{overpic}[width=\linewidth]{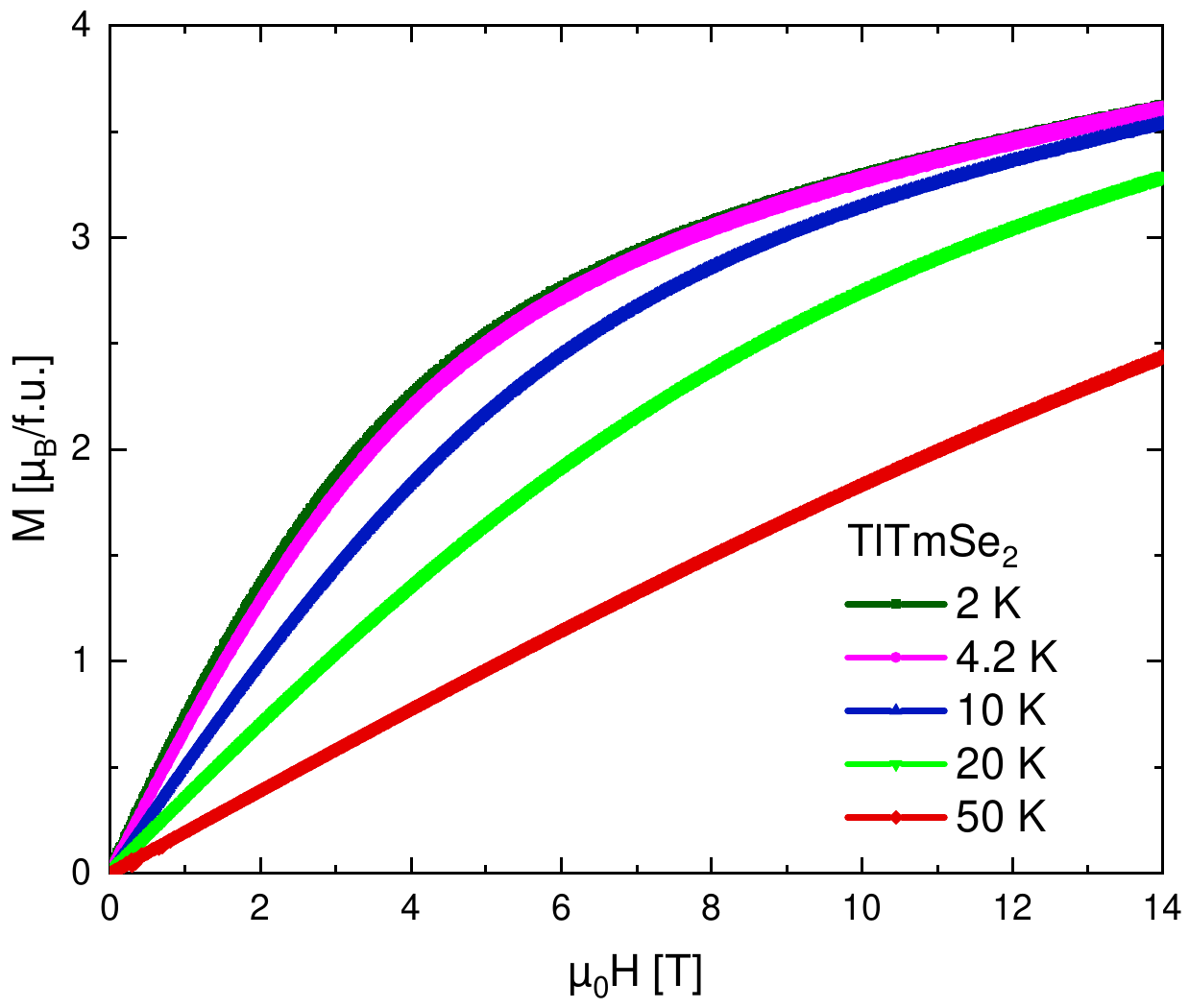}
                \put(1,80){\textbf{b)}}
            \end{overpic}
            \caption{a) Field dependence of the magnetic moment per formula unit of TlErSe$_2$ a) and TlTmSe$_2$ b) for various temperatures.}
            \label{fig:TRES_MxH}
        \end{figure}
        
        Fig. \ref{fig:TRES_MxT} a) shows the magnetization of TlErSe$_2$ divided by the applied field ($M/H$) as a function of temperature in an external magnetic field of $1$\,T. The data was fitted with the Curie-Weiss law, i.e., $M/H(T) = C/(T-\Theta_{CW})$, for temperatures above $200$\,K. The corresponding fitting yields an effective moment of $\mu_{\mbox{eff}} = 9.6 \pm 0.2\,\mu_B$/Er and a Curie-Weiss temperature of $\Theta_{CW}=-(7\pm1)$\,K. 
        The effective moment is in good agreement with the expected value for a trivalent free $Er$-ion of $\mu_{\mbox{eff}} = 9.58\, \mu_B$ \cite{Ashcroft76}. Similar results have been found in other Er-based delafossites such as NaErSe$_2$ and KErSe$_2$, where a $\mu_{\mbox{eff}} = 9.5\,\mu_B$ was reported for both compounds \cite{xing2019synthesis}. The negative $\Theta_{CW}$ indicates sizable antiferromagnetic correlations between the Er$^{3+}$ ions. Down to the lowest measured temperature of $1.8$\,K no sign of ordering is observed, resulting in a minimal frustration factor $f=|\Theta_{CW}/T_N|$ \cite{PhysRevLett.64.2070} of $\approx 3.9$. The insert shows $M/H$ as a function of the temperature for applied fields $\geq2~$T and its first derivative. $dM/dT$ reveals a minimum at $3.2~$K and $2~$T, that shifts towards higher temperatures and broadens with increasing applied field. This minimum is taken as an indication of the gradual field polarization (see Fig. \ref{fig:phase-TES}). This observation agrees well with the field-dependent measurement of the magnetization (Fig. \ref{fig:TRES_MxH} a). At low temperatures, the moment first steeply increases with an increasing field and then slowly tapers off becoming almost linear for applied fields greater than $\sim10$\,T, finally reaching $\sim 6.5~\mu_B$/Er at $14~$T and $2~$K. 

        Fig. \ref{fig:TRES_MxT} b) shows the temperature dependence of the magnetization divided by the applied magnetic field for TlTmSe$_2$. The data show no evidence of a magnetic transition in TlTmSe$_2$ down to $1.8~$K. $M/H$ follows a Curie-Weiss law above $200~$K. The fit yields an effective moment of $\mu_{eff}=(7.5\pm0.1)~\mu_B$/Tm and a Curie-Weiss temperature of $\Theta_{CW} = (-10\pm2)~$K. The obtained effective moment is in good agreement with the theoretical moment of free Tm$^{3+}$ ion (7.57 $\mu_B$) \cite{Ashcroft76}. $\Theta_{CW}$ indicates sizable antiferromagnetic correlations. The resulting frustration factor is at least $f\geq 5$. At around $\sim 100~$K the magnetic behavior starts to deviate from the Curie-Weiss law with the observed magnetization being smaller than the free ion value, indicating a reduction of the Curie constant $C$ as a function of temperature.

        Isothermal magnetization data of TlTmSe$_2$ collected at various temperatures are plotted in Fig. \ref{fig:TRES_MxH} b). The low-temperature data shows a steep increase of the magnetic moment over the first three Teslas up to $\sim 2 ~\mu_B$/f.u., which then slowly transitions into a linear behavior for fields greater than $\sim12~$T.  
        The linear behavior above $\sim 12~$T may at least partly be attributed to a temperature-independent van-Vleck contribution stemming from excitations to higher CEF level.

        It should be noted, however, that the measured samples are polycrystalline and thus the observed finite linear slope at high field may originate not purely from a van-Vleck contribution but from the polycrystalline average. Rare-earth magnets in general, but also more specifically rare-earth-based delafossites commonly display a large $g$-factor anisotropy \cite{Schmidt2021Ybdelafossites, 10.21468/SciPostPhys.9.3.041,scheie2020crystal,xing2019synthesis}. The presented data is a powder average and thus an admixture of the magnetic easy and hard directions. The saturation field of the latter might exceed our experimental accessible range and hence results in the observation of a finite slope for large applied fields. Thus, we refrain from fitting the field-dependent magnetization data.
        
        \begin{figure}
            \centering
            \begin{minipage}{0.49\textwidth}
                \includegraphics[width=\textwidth]{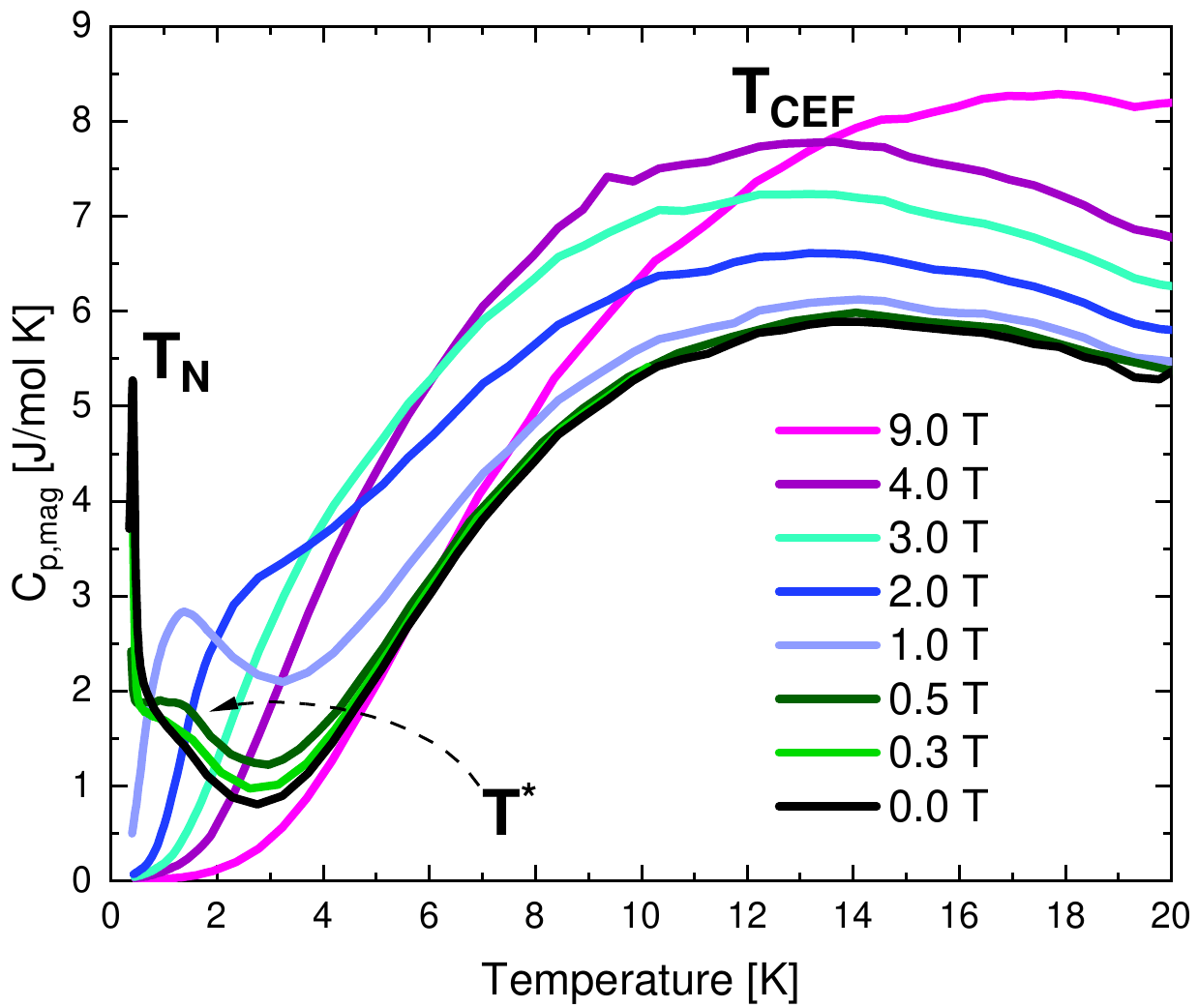}
            \end{minipage}
            \caption{Magnetic contribution of the specific heat $C_{p,mag}$ as a function of temperature for various magnetic fields. $T_N$ marks the transition temperature and $T^*$ corresponds to the shoulder-like feature in addition to the broad maximum $T_{CEF}$ from populated higher CEF level.}
            \label{fig:TES-Cp}
        \end{figure}
        
        \begin{figure}
            \centering
            \includegraphics[width=\linewidth]{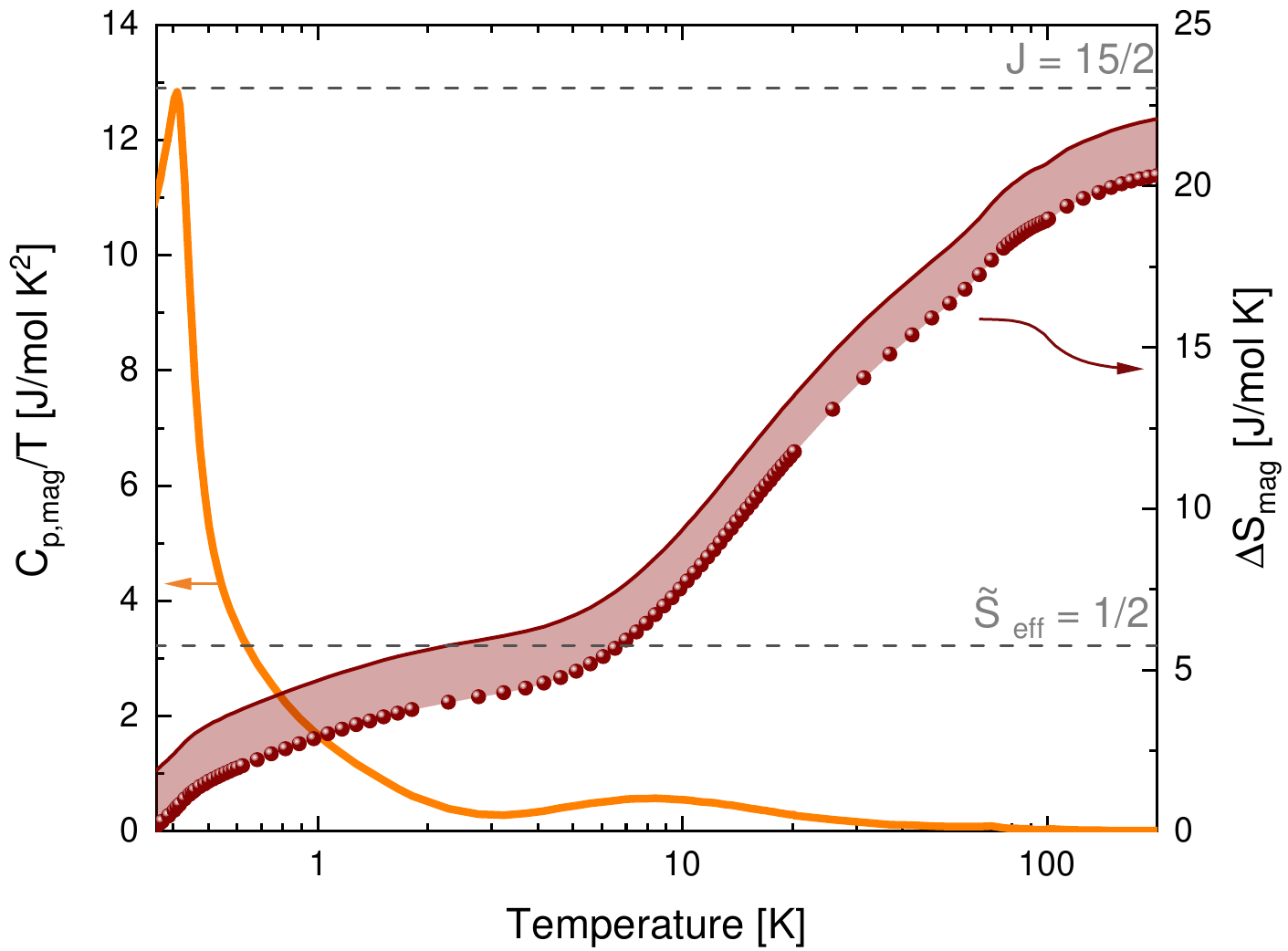}

            \caption{Magnetic contribution of the specific heat divided by temperature $C_{p,mag}/T$ of the zero field measurement (orange solid line, left $y$-axis) and the corresponding integrated magnetic entropy (red dots, right $y$-axis). The solid red line indicates an estimate of the entropy which is missing from the temperature range below $0.35$\,K. The two dashed gray lines mark the expected entropy values for a $ \tilde{S}=1/2$ and $J= 15/2$ state.}
            \label{fig:TES-S}
        \end{figure}
    
        To further explore the low-temperature magnetism of the two samples we extracted the magnetic contribution of the specific heat for various applied fields up to 9\,T measured down to 0.36 mK. The phononic contribution was subtracted by Lindemann-scaling the specific heat of the non-magnetic analogue TlLuSe$_2$. 
        
        $C_{p,mag}$ of TlErSe$_2$ (Fig. \ref{fig:TES-Cp}) reveals a sharp peak at $420$\,mK and zero field. This peak is suppressed by applying an external magnetic field until no sharp specific heat peak is observable down to $0.36$\,K at $0.5$\,T. It marks the onset of long-range antiferromagnetic order. Besides the sharp main peak, a faint shoulder is observed, which develops into a broad hump (labeled $T^*$), shifting toward higher temperatures with increasing fields. This broad feature is typical for quasi 2D systems and indicates the onset of 2D correlations. In addition, a second hump at higher temperatures, i.e., around $T\sim 12$\,K ($T_{\text{CEF}}$) is observed. This broad feature originates from the population of higher CEF levels. $T_{\text{CEF}}$ remains rather stationary for fields up to $4$\,T. In this field range both humps merge and become indistinguishable. At an applied field of $9~$T the combined maximum is shifted to $\sim 17~$K. 
        
        The corresponding entropy release $\Delta S_{mag}$ is presented in Fig. \ref{fig:TES-S} and shows a steep increase corresponding to the sharp specific heat peak of the magnetic phase transition followed by a leveling-off in the low-temperature regime ($T < 6$\,K). The plateau-like feature is close to $R\cdot \ln{(2)}$. This is the expected value for a $ \tilde{S}=1/2$ spin system verifying the ground-state Kramers doublet. With the transition temperature $T_N = 0.420$\,K close to the system's limitation it is not possible to capture the full peak. Therefore some of the entropy at low temperatures is not accounted for when integrating over $C_{p,mag}/T$. A rough estimate of this contribution to $S_{mag}$ is given by the red-shaded area above the data points. $C_{p,mag}$ was fitted with  $C_{p,mag} \propto \ T^3$, which expected for an antiferromagnet close to $T_N$ \cite{Gopal}. Above $6~$K the released entropy increases steadily approaching $R\cdot \ln{(2(15/2)+1)}$ the value of expected for a free Er$^{3+}$ ion.

        \begin{figure}
            \centering
            \includegraphics[width=0.49\textwidth]{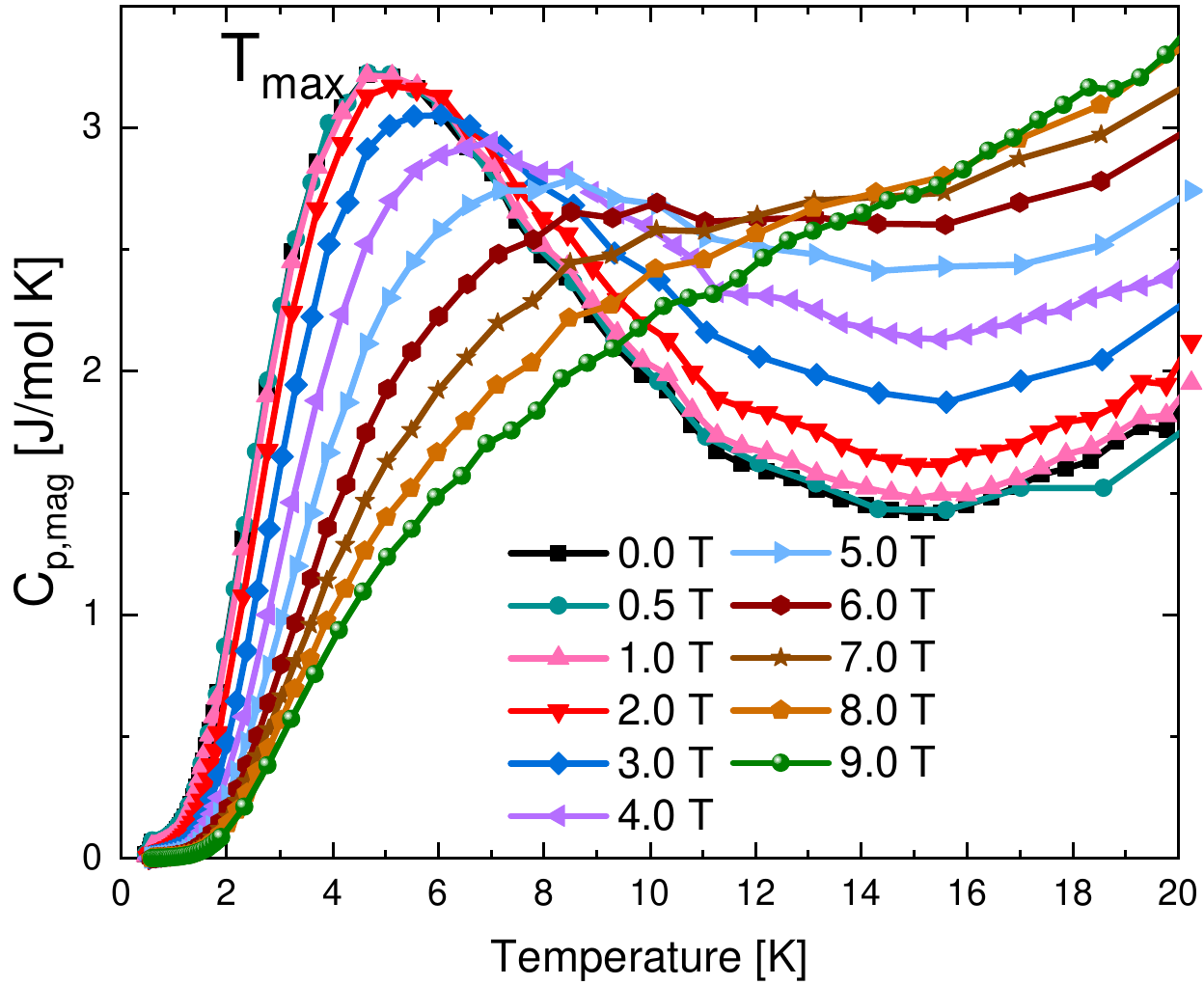}
            \caption{Magnetic contribution of the specific heat $C_{p,mag}$ as a function of temperature of TlTmSe$_2$ for various magnetic fields. $T_{max}$ marks the observed broad maximum which broadens even further for large applied fields.}
            \label{fig:TTS-Cp}
        \end{figure}
        
        \begin{figure}
            \centering
            \includegraphics[width=0.49\textwidth]{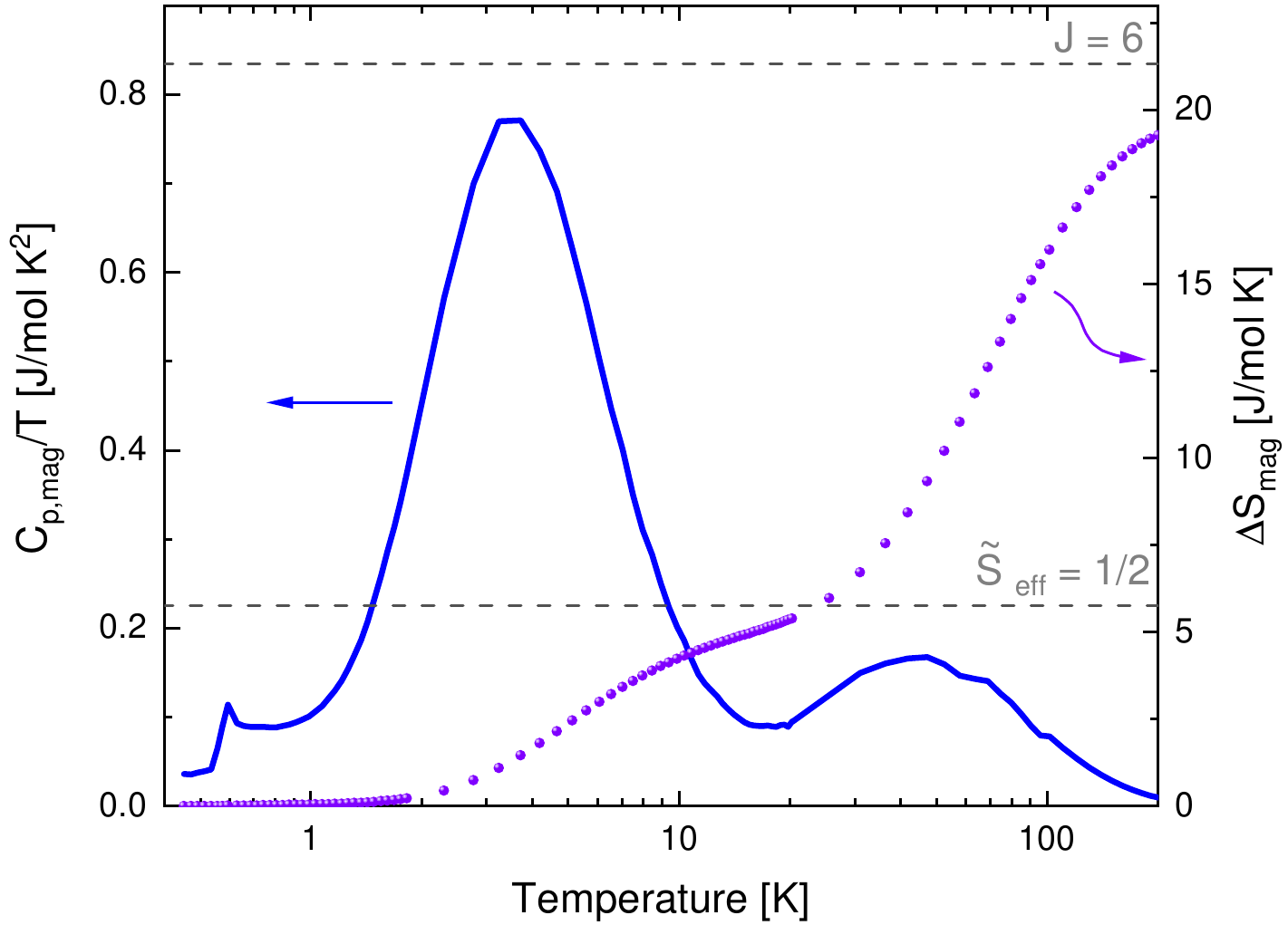}
            \caption{Magnetic contribution of the specific heat divided by temperature $C_{p,mag}/T$ of the zero field measurement (blue solid line, left $y$-axis) and the corresponding integrated magnetic entropy released (purple dots, right $y$-axis). The two dashed gray lines mark the expected entropy values for an effective spin $\tilde{S}= 1/2$ and $J= 6$ state. }
            \label{fig:TTS-S}
        \end{figure}

        \begin{figure}
             \centering
            \includegraphics[width=0.49\textwidth]{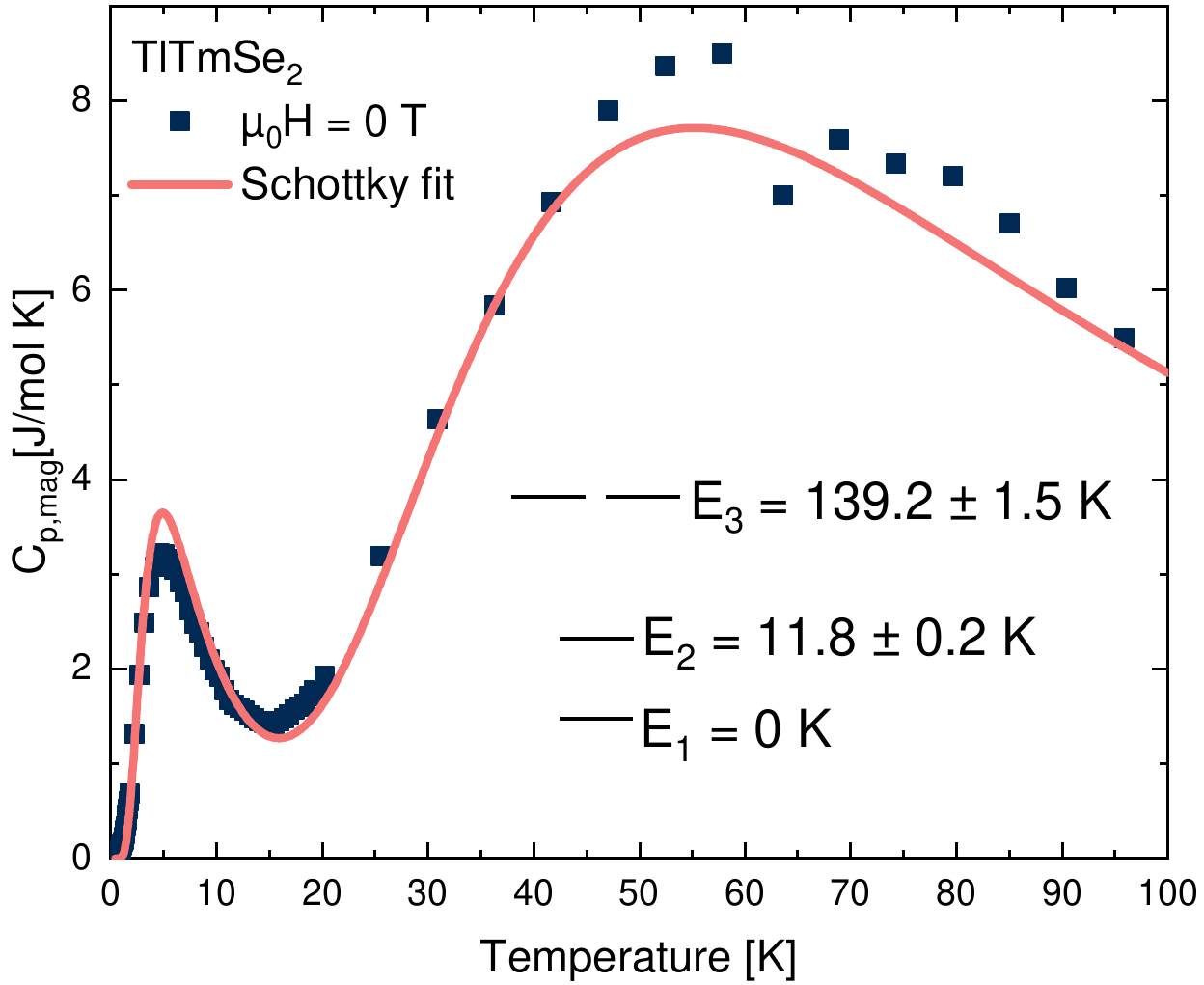}
            \caption{Magnetic contribution of the specific heat of TlTmSe$_2$ as a function of temperature for zero field. The solid line corresponds to a Schottky fit considering three energy levels \cite{Gopal}. The level scheme is depicted and the data was best described with a degeneracy ratio of $1:1:2$, for the energy levels $E_1$, $E_2$, and $E_3$, respectively.}
            \label{fig:schottky-TTS}
        \end{figure}

        The magnetic contribution of the specific heat measurement of TlTmSe$_2$ as a function of temperature is displayed in Fig. \ref{fig:TTS-Cp} for various applied magnetic fields. Throughout the whole temperature range, no sharp anomaly, apart from a small measurement artifact around $\sim0.55~$K \cite{artifact}, is observed suggesting the absence of a spontaneous long-range magnetic order phase transition down to $\sim 0.5~$K. The magnetic contribution of the specific heat exhibits a broad maximum at $T_{max}=5~$K for magnetic fields below $2~$T. With increasing field strength the maximum shifts linearly towards higher temperatures and smears out further. Fig. \ref{fig:TTS-S} shows $C_{p,mag}/T$ at zero-field up to $\sim 200~$K. In the higher temperature regime, another broad feature $T_{CEF}$ is observed at approximately $45~$K. The origin of this bump can be related to the higher crystal electric field levels being populated, which is supported by the released entropy in this temperature regime. The obtained magnetic entropy is shown in the same figure. It experiences a plateau-like feature around $R\cdot \ln{(2)}$ similar to TlErSe$_2$. While this may indicate a $ \tilde{S}=1/2$ regime below $\sim 20~$K, the entropy released for a Schottky-like anomaly originating from two singlet states is expected to yield the same value, as $S$ depends solely on the ratio of degeneracies $g_1/g_0$ for a two-level system \cite{Gopal}. Considering the reduction of $\mu_{eff}$ for low temperatures as well as the CEF calculations presented below the lowest energy levels are (nonmagnetic) singlet states and thus the second scenario is more plausible for the Tm compound. For elevated temperatures, the entropy approaches the expected value for the $J=6$ $4f$ manifold of a free Tm$^{3+}$.

        In Fig. \ref{fig:schottky-TTS} the magnetic contribution of the specific heat was fitted with a Schottky approach using three energy levels $E_1$, $E_2$, and $E_3$, while the first two correspond to the ground and first excited CEF levels the third accounts for higher CEF levels. For the degeneracy $g_i$ of the levels the best fit was obtained with $E_1$ and $E_2$ having the same value, and $g_3$ being twice this value. From this we conclude that the two lowest energy levels are most likely singlet states with an energy gap of $11.8\pm0.2~$K, which is in agreement with the quantum chemical calculations presented below.

\begin{figure}[ht]
        	\centering
            \begin{overpic}[width=\linewidth]{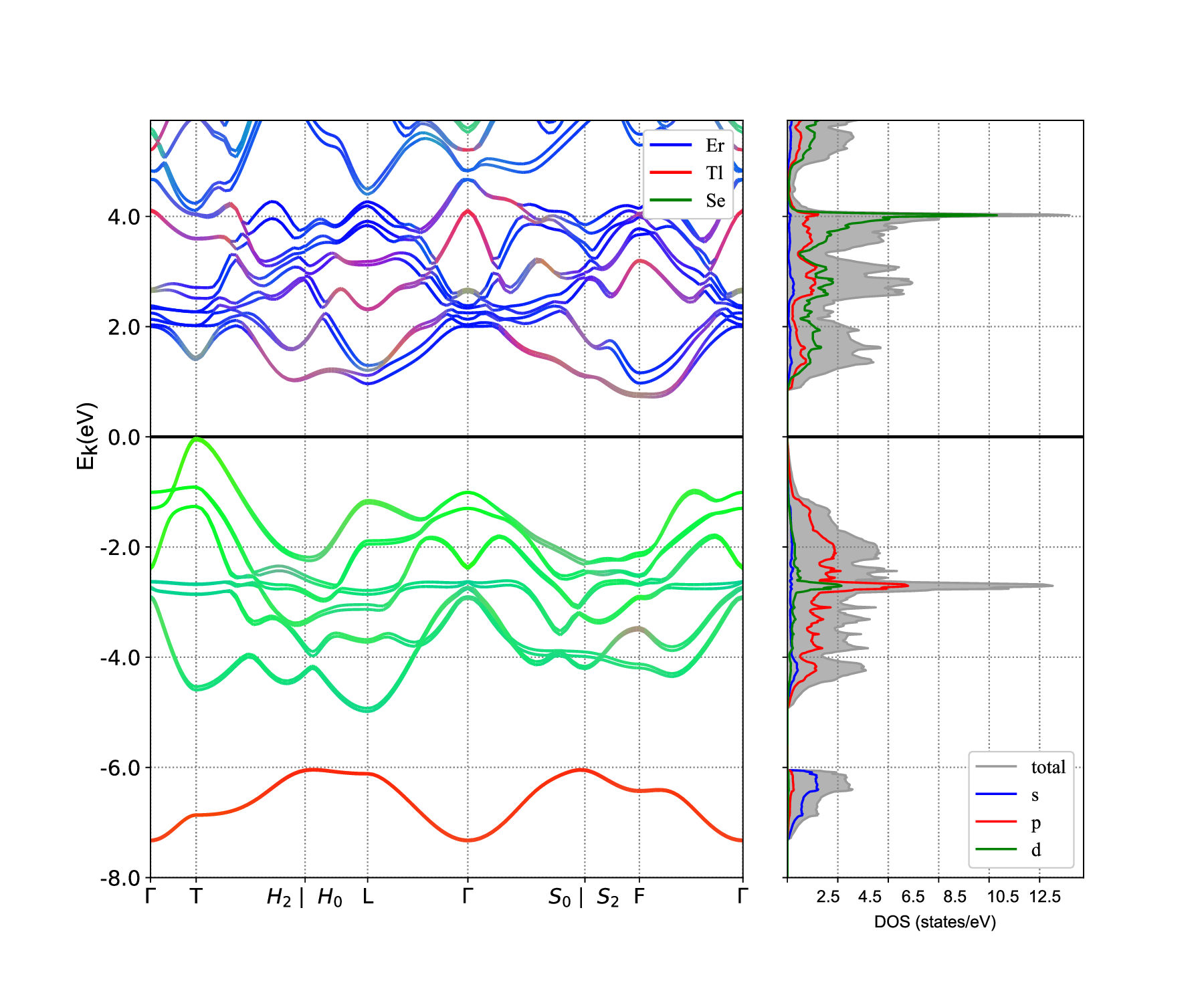}
                \put(5,72){\textbf{a)}}
            \end{overpic}
            \begin{overpic}[width=\linewidth]{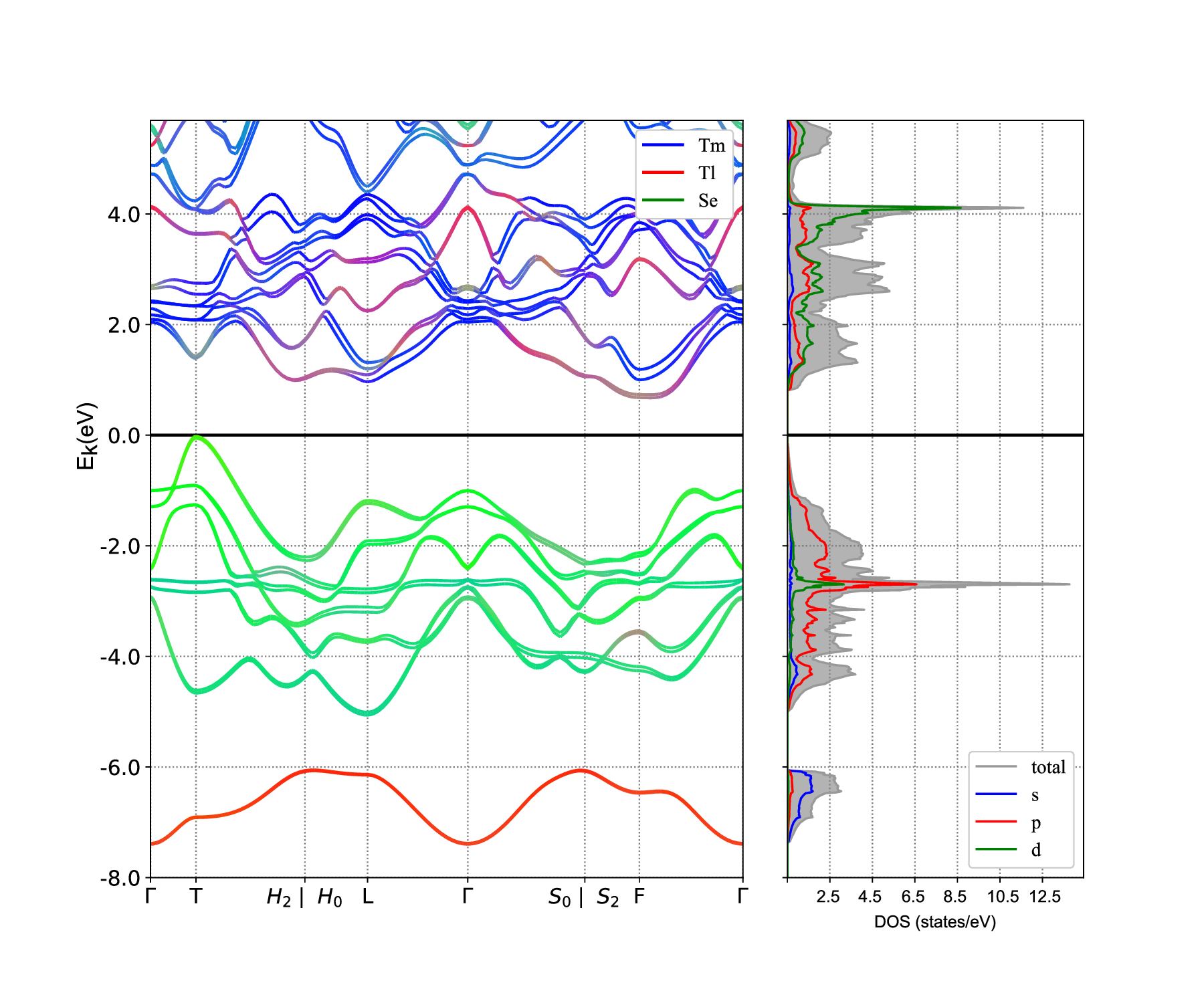}
                \put(5,72){\textbf{b)}}
            \end{overpic}
        	\caption{Band structure and density of states of TlErSe$_2$ a) and TlTMSe$_2$ b).
            \label{fig:band} }
        \end{figure}

        In order to understand the different behavior of the two compounds, we performed DFT and quantum chemical calculations. In the DFT calculation both materials are insulators. Fig. \ref{fig:band} a) and b) show the band structures and density of states (DOS) in TlErSe$_2$ and TlTmSe$_2$ respectively.
        The calculated band structures show strong similarities, with almost identical dispersions over the Brillouin zone for both compounds. This is expected as both compounds are isostructural, their crystal structures are similar and their 4$f$ electrons can be treated as core electrons. 
        
        The top of the valence bands in both materials consists predominantly of $p$-states of Se. The conduction band electrons originate from both the $d$-electron orbitals of Er(Tm) and the $p$-electrons of Tl. The resulting indirect gap is about $0.74~$eV (TlErSe$_2$) and $0.69~$eV (TlTmSe$_2$). This structure ensures that the magnetic response comes entirely from localised $4f$ electrons.

        For insights into the $4f$ electronic structures we carried out quantum chemical calculations. The low-energy NEVPT2 spectrum of TlErSe$_2$ is defined by a group of eight SO doublets with relative energies of 0, 6.5, 7.4, 8.7, 11.3, 30.4, 32.4, and 33.4 meV, related to the $J\!=\!15/2$ free-ion ground-state term. Other excited SO states require excitation energies of at least 0.75 eV. Notably, the NEVPT2 treatment brings significant corrections to the CASSCF relative energies, an aspect that is detailed in the Appendix (Tab. \ref{tab:excited_Er}). The lowest excited state lies somewhat lower compared to the case of 4$f^{13}$ ions in similar crystalline environment \cite{NaYbSe2,Zhang2021NaYbSe2}.

        The low-energy MRCI spectrum for the TlTmSe$_2$ compound is defined by a group of five SO singlets and four SO doublets with relative energies of 0, 0.8, 3.5, 5.1, 9.0, 12.9, 18.7, 19.1, and 21.7 meV, related to the $J\!=\!6$ free-ion ground-state term. Other excited SO states are located above 0.95 eV. The MRCI treatment brings significant corrections to the CASSCF excitation energies, as presented in Table \ref{tab:excited_TlTm}. In other 4$f^{12}$ triangular-lattice compounds, e.g., KTmSe$_2$ \cite{expt_KTm} and NaTmTe$_2$ \cite{NaTmSe2}, the  experimentally determined first excited state is at 1.2 and 2.9 meV, respectively.

 \section{Discussion and Summary}
 
    \begin{figure}
         \centering
        \includegraphics[width=0.49\textwidth]{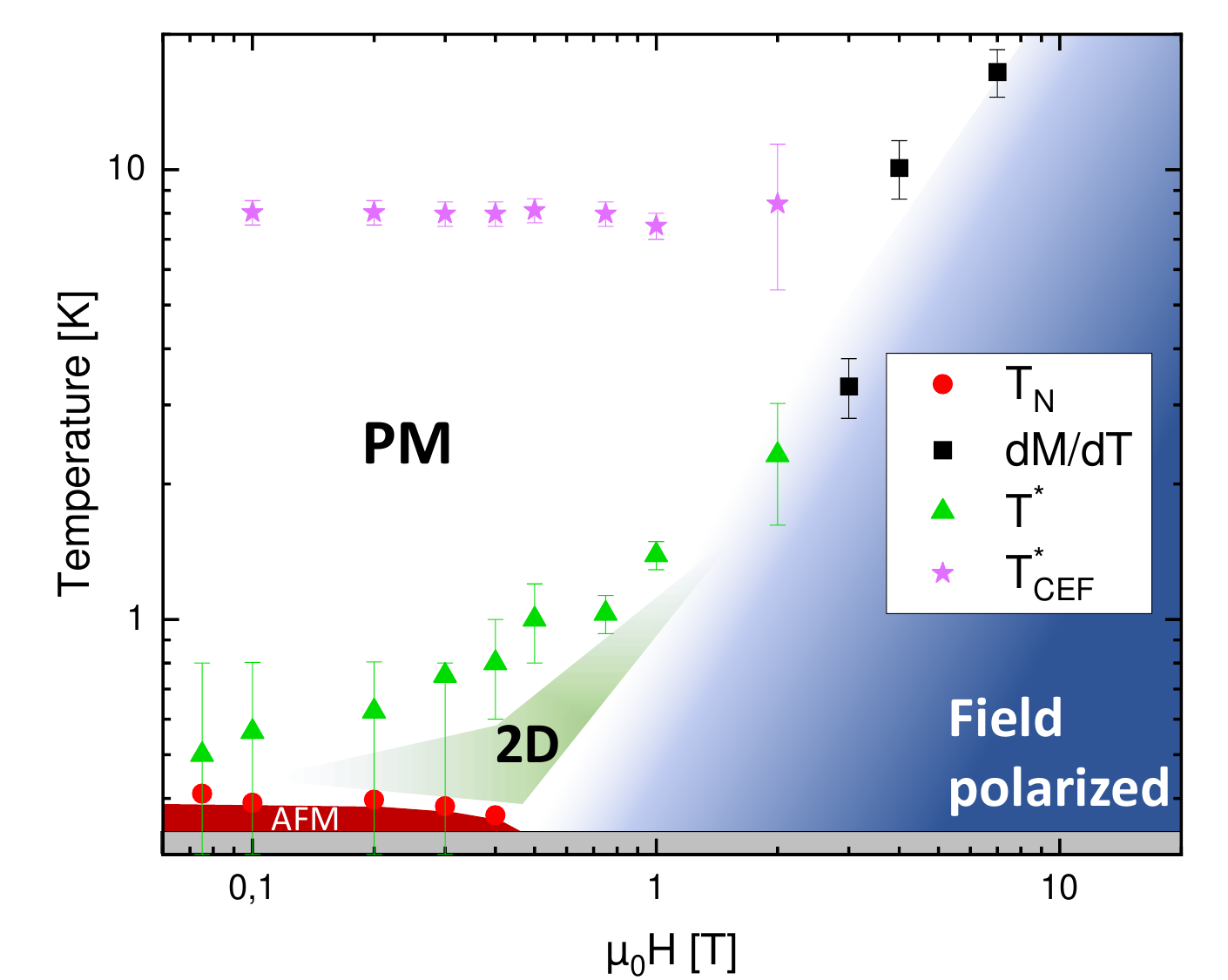}
        \caption{Field-temperature phase diagram of TlErSe$_2$ on a Log--Log scale. Marked are T$_N$ (red circles) corresponding to the sharp peak and the shoulder (light green triangles) which develops into a broad bump (pink stars) observed in $C_{p,mag}$ as well as the onset of saturation as measured by high field susceptibility (black squares); for more information see text. The colored areas are meant as guides-to-the-eye. The grey shaded area below marks the temperature limit of the device.}
        \label{fig:phase-TES}
    \end{figure}

Our results show that TlErSe$_2$ is an example of a magnetically long-range ordered triangular lattice antiferromagnet govern by its spin $ \tilde{S}=1/2$ Kramers doublet ground-state, similar to NaErSe$_2$ \cite{xing2019synthesis}, NaErS$_2$ \cite{Gao2020NaErS}, KCeS$_2$ \cite{10.21468/SciPostPhys.9.3.041,kulbakov2021stripe}, KErSe$_2$\cite{xing2019synthesis,xing2021stripe} or KYbSe$_2$ \cite{scheie2023witnessing}, contrasting putative quantum spin liquid candidates like NaYbSe$_2$ \cite{liu2018rare,Ranjith2019NaYbSe2}, NaYbO$_2$ \cite{Wu2022NaYbO2}, KYbS$_2$ \cite{iizuka2020single} or CsYbSe$_2$ \cite{Xing2019CsRESe2}. While both kinds of behavior are observed within the delafossite family a clear tuning parameter remains unclear and under debate. However, it is evident that anisotropic exchange plays a significant role. For similar Er$^{3+}$ based compounds both easy plane (XY) anisotropy (ErMgGaO$_4$ \cite{PhysRevB.101.094432}, K$_3$Er(VO$_4$)$_2$ \cite{PhysRevB.102.104423}, Er$_2$Ti$_2$O$_7$ , NaErSe$_2$ \cite{xing2019synthesis}, CsErSe$_2$ \cite{scheie2020crystal}) and Ising anisotropy (Ba$_3$ErB$_9$O$_{18}$\cite{PhysRevB.106.104408}) were reported, motivating follow-up studies of single crystalline samples.

Nonetheless, given the temperature range and the large total angular moment of $J=\pm15/2$ associated with the Er$^{3+}$ ground state doublet, here, dipolar interactions can be a driving factor for the onset of long-range magnetic order. Following this the most likely ordering type would be a stripe order with the moments pointing along the $Er-Er$ bond axis.

In addition to the pronounced peak at $T_N$, a shoulder-like feature was observed evolving into a broad hump at larger fields (see Fig. \ref{fig:phase-TES}). Due to the delafossites crystal structure, i.e., stacked rare-earth ion triangular layers well separated by $A^{+}$ spacer layers, the hierarchy of the magnetic couplings ($\mathcal{J}$) closely resembles quasi-2D systems with $\mathcal{J}_{intralayer} >> \mathcal{J}_{interlayer}$. Here, these broad features appearing just before a magnetically ordered state forms correspond to 2D correlations \cite{sengupta2003specific}.

While the work by Liu et al. \cite{liu2018selective} proposed numerous ordered states for non-Kramers doublets on a triangular lattice, no sign indicating long-range magnetic order was found in TlTmSe$_2$ down to $360~$mK. These findings are in line with reports on related compounds such as KTmSe$_2$ \cite{expt_KTm}, NaTmSe$_2$ \cite{LiNaTmS2} or LiTmSe$_2$\cite{LiNaTmS2}. Our quantum chemical calculations suggest a sizable energy gap of $0.8$ or $1.2$ meV between the two lowest singlet states, CASSCF or MRCI respectively. In agreement with this, fitting the low temperature specific heat with a two level Schottky system yields an energy gap and degeneracy ratio of the two lowest level of $11.8~$K and $1:1$, respectively. Following, it is likely, that the absence of a sharp ordering transition in TlTmSe$_2$ can be attributed to the non-magnetic nature of it's ground state singlet and the comparably large energy gap to the next higher CEF level.

In summary, we report the synthesis of high quality polycrystalline Tl$M$Se$_2$ ($M$: Er and Tm) samples. While the erbium compound shows clear signs of the presence of a long-range antiferromagnetic ordered phase below $420~$mK, no such signs were found for TlTmSe$_2$. This profound difference in magnetic properties is attributed to the distinct difference in the CEF-level structure. For TlErSe$_2$ the $ \tilde{S}=1/2$ kramer doublet ground state of the Er$^{3+}$ ions most likely is dominated by contributions of large $\ket{m_J}$ which results in a sizable dipolar coupling, strong enough to stabilize long-range order in the $^3$He temperature range. With this coupling mechanism not present in the well separated non-magnetic singlet ground state of the Tm$^{3+}$ ions, here, no long-range order phase can form.

\section{Acknowledgements}
We acknowledge financial support from the German Research Foundation (DFG) within the SFB 1143 Correlated Magnetism From Frustration to Topology, project-id 247310070 (Projects B01, B03 and A05) and project No.390858490 (W\"urzburg-Dresden Cluster of Excellence on Complexity and Topology in Quantum Matter – ct.qmat, EXC 2147)
We appreciate the ICP-OES measurements by Dr. Gudrun Auffermann (MPI-CPfS Dresden).
D.V.E. and A.N. thank M. Richter for discussions and U. Nitzsche for technical support and acknowledge financial support from the DFG, projects 529677299, 455319354.
P.B. and L.H. thank T. Petersen for discussions, U. Nitzsche for technical support, and the DFG (Project No. 441216021) for financial support.

\bibliographystyle{apsrev4-2}
\bibliography{bib}

@article{Xing2019CsRESe2,
  doi = {10.1021/acsmaterialslett.9b00464},
  url = {https://doi.org/10.1021/acsmaterialslett.9b00464},
  year = {2019},
  month = nov,
  publisher = {American Chemical Society ({ACS})},
  volume = {2},
  number = {1},
  pages = {71--75},
  author = {Jie Xing and Liurukara D. Sanjeewa and Jungsoo Kim and G. R. Stewart and Mao-Hua Du and Fernando A. Reboredo and Radu Custelcean and Athena S. Sefat},
  title = {Crystal Synthesis and Frustrated Magnetism in Triangular Lattice CsRESe2 (RE = La−Lu): Quantum Spin Liquid Candidates CsCeSe2 and CsYbSe2},
  journal = {{ACS} Mater. Lett.}
}

@article{Wu2022NaYbO2,
  url = {https://doi.org/10.1007/s44214-022-00011-z},
  year = {2022},
  month = nov,
  publisher = {Springer Science and Business Media {LLC}},
  volume = {1},
  pages = {13},
  number = {1},
  author = {Jiangtao Wu and Jianshu Li and Zheng Zhang and Changle Liu and Yong Hao Gao and Erxi Feng and Guochu Deng and Qingyong Ren and Zhe Wang and Rui Chen and Jan Embs and Fengfeng Zhu and Qing Huang and Ziji Xiang and Lu Chen and Yan Wu and E. S. Choi and Zhe Qu and Lu Li and Junfeng Wang and Haidong Zhou and Yixi Su and Xiaoqun Wang and Gang Chen and Qingming Zhang and Jie Ma},
  title = {Magnetic field effects on the quantum spin liquid behaviors of NaYbS$_2$},
  journal = {Quantum Frontiers},
  doi = {10.1007/s44214-022-00011-z}
}

@article{Ferreira2020,
  year = {2020},
  month = feb,
  publisher = {Frontiers Media {SA}},
  volume = {8},
 pages = {127},
  author = {Timothy Ferreira and Jie Xing and Liurukara D. Sanjeewa and Athena S. Sefat},
  title = {Frustrated Magnetism in Triangular Lattice TlYbS$_2$ Crystals Grown via Molten Flux},
 journal = {Front. Chem.},
 doi = {10.3389/fchem.2020.00127}
}

@article{Gao2020NaErS,
  doi = {10.1103/physrevb.102.024424},
  url = {https://doi.org/10.1103/physrevb.102.024424}   ,
  year = {2020},
  pages = {024424},
  month = jul,
  publisher = {American Physical Society ({APS})},
  volume = {102},
  number = {2},
  author = {Shang Gao and Fan Xiao and Kazuya Kamazawa and Kazuhiko Ikeuchi and Daniel Biner and Karl W. Kr\"{a}mer and Christian R\"{u}egg and Taka-hisa Arima},
  title = {Crystal electric field excitations in the quantum spin liquid candidate NaErS$_2$},
  journal = {Phys. Rev. B}
}

@article{Baenitz2018,
  title = {NaYbS$_2$: A planar spin-$\frac{1}{2}$ triangular-lattice magnet and putative spin liquid},
  author = {Baenitz, M. and Schlender, Ph. and Sichelschmidt, J. and Onykiienko, Y. A. and Zangeneh, Z. and Ranjith, K. M. and Sarkar, R. and Hozoi, L. and Walker, H. C. and Orain, J.-C. and Yasuoka, H. and van den Brink, J. and Klauss, H. H. and Inosov, D. S. and Doert, Th.},
  journal = {Phys. Rev. B},
  volume = {98},
  issue = {22},
  pages = {220409},
  numpages = {6},
  year = {2018},
  month = {Dec},
  publisher = {American Physical Society},
  doi = {10.1103/PhysRevB.98.220409},
  url = {https://link.aps.org/doi/10.1103/PhysRevB.98.220409}
}

@Article{Schmidt2021Ybdelafossites,
  author    = {Schmidt, B. and Sichelschmidt, J. and Ranjith, K. M. and Doert, Th. and Baenitz, M.},
  title     = {Yb delafossites: Unique exchange frustration of $4f$ spin-$\frac{1}{2}$ moments on a perfect triangular lattice},
  journal   = {Phys. Rev. B},
  year      = {2021},
  volume    = {103},
  issue     = {21},
  month     = {Jun},
  pages     = {214445},
  doi       = {10.1103/Schmidt2021Ybdelafossites},
  url       = {https://link.aps.org/doi/10.1103/Schmidt2021Ybdelafossites},
  numpages  = {18},
  publisher = {American Physical Society},
}

@Article{Ranjith2019NaYbO2,
  author    = {Ranjith, K. M. and Dmytriieva, D. and Khim, S. and Sichelschmidt, J. and Luther, S. and Ehlers, D. and Yasuoka, H. and Wosnitza, J. and Tsirlin, A. A. and K\"uhne, H. and Baenitz, M.},
  title     = {Field-induced instability of the quantum spin liquid ground state in the ${J}_{\mathrm{eff}}=\frac{1}{2}$ triangular-lattice compound NaYbO$_2$},
  journal   = {Phys. Rev. B},
  year      = {2019},
  volume    = {99},
  issue     = {18},
  month     = {May},
  pages     = {180401},
  doi       = {10.1103/PhysRevB.99.180401},
  url       = {https://link.aps.org/doi/10.1103/PhysRevB.99.180401},
  numpages  = {7},
  publisher = {American Physical Society},
}

@Article{Sichelschmidt_2019,
  author    = {J\"{o}rg Sichelschmidt and Philipp Schlender and Burkhard Schmidt and Michael Baenitz and Thomas Doert},
  title     = {Electron spin resonance on the spin-1/2 triangular magnet NaYbS$_2$},
  journal   = {J. Condens. Matter Phys.},
  year      = {2019},
  volume    = {31},
  number    = {20},
  month     = {mar},
  pages     = {205601},
  doi       = {10.1088/1361-648x/ab071d },
  publisher = {{IOP} Publishing},
}

@Article{Ranjith2019NaYbSe2,
  author    = {Ranjith, K. M. and Luther, S. and Reimann, T. and Schmidt, B. and Schlender, Ph. and Sichelschmidt, J. and Yasuoka, H. and Strydom, A. M. and Skourski, Y. and Wosnitza, J. and K\"uhne, H. and Doert, Th. and Baenitz, M.},
  title     = {Anisotropic field-induced ordering in the triangular-lattice quantum spin liquid ${\mathrm{NaYbSe}}_{2}$},
  journal   = {Phys. Rev. B},
  year      = {2019},
  volume    = {100},
  issue     = {22},
  month     = {Dec},
  pages     = {224417},
  doi       = {10.1103/PhysRevB.100.224417},
  url       = {https://link.aps.org/doi/10.1103/PhysRevB.100.224417},
  numpages  = {11},
  publisher = {American Physical Society},
}

@Article{Ding2019NaYbO2,
  author    = {Ding, Lei and Manuel, Pascal and Bachus, Sebastian and Gru\ss{}ler, Franziska and Gegenwart, Philipp and Singleton, John and Johnson, Roger D. and Walker, Helen C. and Adroja, Devashibhai T. and Hillier, Adrian D. and Tsirlin, Alexander A.},
  title     = {Gapless spin-liquid state in the structurally disorder-free triangular antiferromagnet ${\mathrm{NaYbO}}_{2}$},
  journal   = {Phys. Rev. B},
  year      = {2019},
  volume    = {100},
  issue     = {14},
  month     = {Oct},
  pages     = {144432},
  doi       = {10.1103/PhysRevB.100.144432},
  url       = {https://link.aps.org/doi/10.1103/PhysRevB.100.144432},
  numpages  = {9},
  publisher = {American Physical Society},
}

@Article{xing2019synthesis,
  author    = {Xing, Jie and Sanjeewa, Liurukara D and Kim, Jungsoo and Meier, William R and May, Andrew F and Zheng, Qiang and Custelcean, Radu and Stewart, GR and Sefat, Athena S},
  title     = {Synthesis, magnetization, and heat capacity of triangular lattice materials NaErSe 2 and KErSe 2},
  journal   = {Physical Review Materials},
  year      = {2019},
  volume    = {3},
  number    = {11},
  pages     = {114413},
  publisher = {APS},
}

@Article{Zhang2021NaYbSe2,
  author    = {Zhang, Zheng and Ma, Xiaoli and Li, Jianshu and Wang, Guohua and Adroja, D. T. and Perring, T. P. and Liu, Weiwei and Jin, Feng and Ji, Jianting and Wang, Yimeng and Kamiya, Yoshitomo and Wang, Xiaoqun and Ma, Jie and Zhang, Qingming},
  title     = {Crystalline electric field excitations in the quantum spin liquid candidate ${\mathrm{NaYbSe}}_{2}$},
  journal   = {Phys. Rev. B},
  year      = {2021},
  volume    = {103},
  issue     = {3},
  month     = {Jan},
  pages     = {035144},
  doi       = {10.1103/PhysRevB.103.035144},
  url       = {https://link.aps.org/doi/10.1103/PhysRevB.103.035144},
  numpages  = {6},
  publisher = {American Physical Society},
}

@Article{10.21468/SciPostPhys.9.3.041,
  author    = {G. Bastien and B. Rubrecht and E. Haeussler and P. Schlender and Z. Zangeneh and S. Avdoshenko and R. Sarkar and A. Alfonsov and S. Luther and Y. A. Onykiienko and H. C. Walker and H. K\"{u}hne and V. Grinenko and Z. Guguchia and V. Kataev and H. -H. Klauss and L. Hozoi and J. van den Brink and D. S. Inosov and B. B\"{u}chner and A. U. B. Wolter and T. Doert},
  title     = {{Long-range magnetic order in the ${\tilde S}=1/2$ triangular lattice antiferromagnet KCeS$_2$}},
  journal   = {SciPost Phys.},
  year      = {2020},
  volume    = {9},
  pages     = {041},
  doi       = {10.21468/SciPostPhys.9.3.041 },
  url       = {https://scipost.org/10.21468/SciPostPhys.9.3.041 },
  publisher = {SciPost},
}

@Book{Gopal,
  author = {E. S. R. Gopal},
  title  = {Specific heats at low temperatures},
  year   = {1966},
  isbn   = {978-1-4684-9083-1},
}

@Book{Ashcroft76,
  author    = {Ashcroft, N. W. and Mermin, N. D.},
  title     = {{S}olid {S}tate {P}hysics},
  year      = {1976},
  publisher = {Holt-Saunders},
}

@Article{Dai2021NaYbSe2,
  author    = {Dai, Peng-Ling and Zhang, Gaoning and Xie, Yaofeng and Duan, Chunruo and Gao, Yonghao and Zhu, Zihao and Feng, Erxi and Tao, Zhen and Huang, Chien-Lung and Cao, Huibo and Podlesnyak, Andrey and Granroth, Garrett E. and Everett, Michelle S. and Neuefeind, Joerg C. and Voneshen, David and Wang, Shun and Tan, Guotai and Morosan, Emilia and Wang, Xia and Lin, Hai-Qing and Shu, Lei and Chen, Gang and Guo, Yanfeng and Lu, Xingye and Dai, Pengcheng},
  title     = {Spinon Fermi Surface Spin Liquid in a Triangular Lattice Antiferromagnet ${\mathrm{NaYbSe}}_{2}$},
  journal   = {Phys. Rev. X},
  year      = {2021},
  volume    = {11},
  issue     = {2},
  month     = {May},
  pages     = {021044},
  doi       = {10.1103/PhysRevX.11.021044 },
  url       = {https://link.aps.org/doi/10.1103/PhysRevX.11.021044},
  numpages  = {10},
  publisher = {American Physical Society},
}

@Article{Bordelon2020NaYbO2,
  author    = {Bordelon, Mitchell M. and Liu, Chunxiao and Posthuma, Lorenzo and Sarte, P. M. and Butch, N. P. and Pajerowski, Daniel M. and Banerjee, Arnab and Balents, Leon and Wilson, Stephen D.},
  title     = {Spin excitations in the frustrated triangular lattice antiferromagnet ${\mathrm{NaYbO}}_{2}$},
  journal   = {Phys. Rev. B},
  year      = {2020},
  volume    = {101},
  issue     = {22},
  month     = {Jun},
  pages     = {224427},
  doi       = {10.1103/PhysRevB.101.224427 },
  url       = {https://link.aps.org/doi/10.1103/PhysRevB.101.224427},
  numpages  = {15},
  publisher = {American Physical Society},
}

@Article{liu2018rare,
  doi = {10.1088/0256-307x/35/11/117501},
  author    = {Liu, Weiwei and Zhang, Zheng and Ji, Jianting and Liu, Yixuan and Li, Jianshu and Wang, Xiaoqun and Lei, Hechang and Chen, Gang and Zhang, Qingming},
  title     = {Rare-earth chalcogenides: A large family of triangular lattice spin liquid candidates},
  journal   = {Chin. Phys. Lett.},
  year      = {2018},
  volume    = {35},
  number    = {11},
  pages     = {117501},
  publisher = {IOP Publishing},
}

@Article{kulbakov2021stripe,
  doi = {https://doi.org/10.1088/1361-648x/ac15d6},
  author    = {Kulbakov, Anton A and Avdoshenko, Stanislav M and Puente-Orench, In{\'e}s and Deeb, Mahmoud and Doerr, Mathias and Schlender, Philipp and Doert, Thomas and Inosov, Dmytro S},
  title     = {Stripe-yz magnetic order in the triangular-lattice antiferromagnet KCeS2},
  journal   = {J. Condens. Matter Phys.},
  year      = {2021},
  volume    = {33},
  number    = {42},
  pages     = {425802},
  publisher = {IOP Publishing},
}

@Article{duczmal1995magnetic,
  author    = {Duczmal, Marek and Pawlak, Lucjan},
  title     = {Magnetic and structural characterization of TlLnSe2 compounds (Ln = Nd to Yb)},
  journal   = {J. Alloys Compd.},
  year      = {1995},
  volume    = {225},
  number    = {1-2},
  pages     = {181--184},
  publisher = {Elsevier},
}

@article{duczmal1994synthesis,
  doi = {10.1016/0925-8388(94)91112-6},
  url = {https://doi.org/10.1016/0925-8388(94)91112-6},
  year = {1994},
  month = jul,
  publisher = {Elsevier {BV}},
  volume = {209},
  number = {1-2},
  pages = {271--274},
  author = {Marek Duczmal and Lucjan Pawlak},
  title = {Magnetic properties of {TiLnS}2 compounds (Ln =Nd, Gd, Dy, Er and Yb)},
  journal = {J. Alloys Compd.}
}

@Book{duczmal2003struktura,
  author    = {Duczmal, Marek},
  title     = {Struktura, w{\l}a{\'s}ciwo{\'s}ci magnetyczne i pole krystaliczne w potr{\'o}jnych chalkogenkach lantanowc{\'o}w i talu TlLnX2 (X},
  year      = {2003},
  publisher = {Oficyna Wydawnicza Politechniki Wroc{\l}awskiej},
}

@Article{kabre1974crystallographic,
  author    = {Kabr{\'e}, S and Julien-Pouzol, M and Guittard, M},
  title     = {Crystallographic study of ternary combinations of thallium and rare-earths with sulfur, selenium of tellurium in binary systems Tl2X-L2X3},
  journal   = {Bulletin de la societ{\'e} chimique de France, Partie I},
  year      = {1974},
  number    = {9-10},
  pages     = {1881--1884},
  publisher = {GAUTHIER-VILLARS 120 BLVD SAINT-GERMAIN, 75280 PARIS, FRANCE},
}

@Article{xing2021stripe,
  author    = {Xing, Jie and Taddei, Keith M and Sanjeewa, Liurukara D and Fishman, Randy S and Daum, Marcus and Mourigal, Martin and dela Cruz, C and Sefat, Athena S},
  title     = {Stripe antiferromagnetic ground state of the ideal triangular lattice compound KErSe2},
  doi = {10.1103/PhysRevB.103.144413},
  journal   = {Phys. Rev. B},
  year      = {2021},
  volume    = {103},
  number    = {14},
  pages     = {144413},
  publisher = {APS},
}

@Article{bordelon2019field,
  doi = {10.1038/s41567-019-0594-5},
  author    = {Bordelon, Mitchell M and Kenney, Eric and Liu, Chunxiao and Hogan, Tom and Posthuma, Lorenzo and Kavand, Marzieh and Lyu, Yuanqi and Sherwin, Mark and Butch, Nicholas P and Brown, Craig and others},
  title     = {Field-tunable quantum disordered ground state in the triangular-lattice antiferromagnet NaYbO2},
  journal   = {Nat. Phys.},
  year      = {2019},
  volume    = {15},
  number    = {10},
  pages     = {1058--1064},
  publisher = {Nature Publishing Group},
}

@InProceedings{iizuka2020single,
  doi = {10.7566/jpscp.30.011097},
  url = {https://doi.org/10.7566/jpscp.30.011097},
  author    = {Iizuka, Ryosuke and Michimura, Shinji and Numakura, Ryosuke and Uwatoko, Yoshiya and Kosaka, Masashi},
  title     = {Single crystal growth and physical properties of ytterbium sulfide KYbS2 with triangular lattice},
  booktitle = {JPS Conf. Proc.},
  year      = {2020},
  pages     = {011097},
}

@Article{PhysRevB.101.094432,
  author    = {Cai, Y. and Lygouras, C. and Thomas, G. and Wilson, M. N. and Beare, J. and Sharma, S. and Marjerrison, C. A. and Yahne, D. R. and Ross, K. A. and Gong, Z. and Uemura, Y. J. and Dabkowska, H. A. and Luke, G. M.},
  title     = {$\ensuremath{\mu}\mathrm{SR}$ study of the triangular Ising antiferromagnet ${\mathrm{ErMgGaO}}_{4}$},
  journal   = {Phys. Rev. B},
  year      = {2020},
  volume    = {101},
  issue     = {9},
  month     = {Mar},
  pages     = {094432},
  doi       = {10.1103/PhysRevB.101.094432},
  url       = {https://link.aps.org/doi/10.1103/PhysRevB.101.094432},
  numpages  = {5},
  publisher = {American Physical Society},
}

@Article{scheie2020crystal,
    doi = {10.1103/PhysRevB.101.144432},
  author    = {Scheie, Allen and Garlea, Vasile O and Sanjeewa, Liurukara D and Xing, Jie and Sefat, Athena Safa},
  title     = {Crystal-field Hamiltonian and anisotropy in KErSe2 and CsErSe2},
  journal   = {Phys. Rev. B},
  year      = {2020},
  volume    = {101},
  number    = {14},
  pages     = {144432},
  publisher = {APS},
}

@misc{scheie2023witnessing,
      title={Witnessing quantum criticality and entanglement in the triangular antiferromagnet {KYbSe}$_2$}, 
      author={A. O. Scheie and E. A. Ghioldi and J. Xing and J. A. M. Paddison and N. E. Sherman and M. Dupont and L. D. Sanjeewa and Sangyun Lee and A. J. Woods and D. Abernathy and D. M. Pajerowski and T. J. Williams and Shang-Shun Zhang and L. O. Manuel and A. E. Trumper and C. D. Pemmaraju and A. S. Sefat and D. S. Parker and T. P. Devereaux and R. Movshovich and J. E. Moore and C. D. Batista and D. A. Tennant},
      year={2023},
      eprint={2109.11527},
      archivePrefix={arXiv},
      primaryClass={cond-mat.str-el}
}

@Article{PhysRevB.106.104408,
  author    = {Khatua, J. and Pregelj, M. and Elghandour, A. and Jagli\ifmmode \check{c}\else \v{c}\fi{}ic, Z. and Klingeler, R. and Zorko, A. and Khuntia, P.},
  title     = {Magnetic properties of the triangular-lattice antiferromagnets ${\mathrm{Ba}}_{3}{R\mathrm{B}}_{9}{\mathrm{O}}_{18}$ $(R=\mathrm{Yb}, \mathrm{Er})$},
  journal   = {Phys. Rev. B},
  year      = {2022},
  volume    = {106},
  issue     = {10},
  month     = {Sep},
  pages     = {104408},
  numpages  = {10},
  publisher = {American Physical Society},
  doi       = {10.1103/PhysRevB.106.104408}
}

@Article{PhysRevB.102.104423,
  author    = {Yahne, Danielle R. and Sanjeewa, Liurukara D. and Sefat, Athena S. and Stadelman, Bradley S. and Kolis, Joseph W. and Calder, Stuart and Ross, Kate A.},
  title     = {Pseudospin versus magnetic dipole moment ordering in the isosceles triangular lattice material ${\mathrm{K}}_{3}{\mathrm{Er}(\mathrm{VO}}_{4}{)}_{2}$},
  journal   = {Phys. Rev. B},
  year      = {2020},
  volume    = {102},
  issue     = {10},
  month     = {Sep},
  pages     = {104423},
  doi       = {10.1103/PhysRevB.102.104423 },
  url       = {https://link.aps.org/doi/10.1103/PhysRevB.102.104423 },
  numpages  = {9},
  publisher = {American Physical Society},
}

@article{Dolg_Stoll_Preuss,
	author = {Dolg,M. and Stoll,H. and Preuss,H. },
	title = {Energy-adjusted ab initio pseudopotentials for the rare earth elements},
	journal = {J. Chem. Phys.},
	volume = {90},
	number = {3},
	pages = {1730-1734},
	year = {1989},
	doi = {10.1063/1.456066},
	URL = {https://doi.org/10.1063/1.456066}
}

@article{Aravena_et_al,
author = {Aravena, Daniel and Neese, Frank and Pantazis, Dimitrios A.},
title = {Improved Segmented All-Electron Relativistically Contracted Basis Sets for the Lanthanides},
journal = {J. Chem. Theory Comput.},
volume = {12},
number = {3},
pages = {1148-1156},
year = {2016},
doi = {10.1021/acs.jctc.5b01048}
}

@article{Dixon_Se,
author = {de Jong,W. A  and Harrison,R. J.  and Dixon,D. A. },
title = {Parallel Douglas–Kroll energy and gradients in NWChem: Estimating scalar relativistic effects using Douglas–Kroll contracted basis sets},
journal = {J. Chem. Phys.},
volume = {114},
number = {1},
pages = {48-53},
year = {2001},
doi = {10.1063/1.1329891}
}

@article{NaYbSe2,
  title = {{Yb}$^{3+}$ $f$-$f$ excitations in {NaYbSe}$_{2}$: Benchmarking embedded-cluster quantum chemical schemes for $4f$ insulators},
  author = {Bhattacharyya, P. and Hozoi, L.},
  journal = {Phys. Rev. B},
  volume = {105},
  issue = {23},
  pages = {235117},
  numpages = {4},
  year = {2022},
  month = {Jun},
  publisher = {American Physical Society},
  doi = {10.1103/PhysRevB.105.235117},
  url = {https://link.aps.org/doi/10.1103/PhysRevB.105.235117}
}

@article{Kresse1999,
  title = {From ultrasoft pseudopotentials to the projector augmented-wave method},
  author = {Kresse, G. and Joubert, D.},
  journal = {Phys. Rev. B},
  volume = {59},
  issue = {3},
  pages = {1758--1775},
  numpages = {0},
  year = {1999},
  month = {Jan},
  publisher = {American Physical Society},
  doi = {10.1103/PhysRevB.59.1758},
  url = {https://link.aps.org/doi/10.1103/PhysRevB.59.1758}
}

@article{Kresse1996,
  title = {Efficient iterative schemes for ab initio total-energy calculations using a plane-wave basis set},
  author = {Kresse, G. and Furthm\"uller, J.},
  journal = {Phys. Rev. B},
  volume = {54},
  issue = {16},
  pages = {11169--11186},
  numpages = {0},
  year = {1996},
  month = {Oct},
  publisher = {American Physical Society},
  doi = {10.1103/PhysRevB.54.11169},
  url = {https://link.aps.org/doi/10.1103/PhysRevB.54.11169}
}

@article{Kresse1993,
  title = {Ab initio molecular dynamics for open-shell transition metals},
  author = {Kresse, G. and Hafner, J.},
  journal = {Phys. Rev. B},
  volume = {48},
  issue = {17},
  pages = {13115--13118},
  numpages = {0},
  year = {1993},
  month = {Nov},
  publisher = {American Physical Society},
  doi = {10.1103/PhysRevB.48.13115},
  url = {https://link.aps.org/doi/10.1103/PhysRevB.48.13115}
}

@article{pymatgen,
  title={Python Materials Genomics (pymatgen): A robust, open-source python library for materials analysis},
  author={Ong, Shyue Ping and Richards, William Davidson and Jain, Anubhav and Hautier, Geoffroy and Kocher, Michael and Cholia, Shreyas and Gunter, Dan and Chevrier, Vincent L and Persson, Kristin A and Ceder, Gerbrand},
  journal={Computational Materials Science},
  volume={68},
  pages={314--319},
  year={2013},
  publisher={Elsevier}
}

@Article{PhysRevLett.64.2070,
  author    = {Ramirez, A. P. and Espinosa, G. P. and Cooper, A. S.},
  title     = {Strong frustration and dilution-enhanced order in a quasi-2D spin glass},
  journal   = {Phys. Rev. Lett.},
  year      = {1990},
  volume    = {64},
  number    = {17},
  issue     = {17},
  pages     = {2070},
  doi       = {10.1103/PhysRevLett.64.2070},
  url       = {https://link.aps.org/doi/10.1103/PhysRevLett.64.2070},
  numpages  = {0},
  publisher = {APS},
}

@article{Shen2019intertwined,
  doi = {10.1038/s41467-019-12410-3},
  url = {https://doi.org/10.1038/s41467-019-12410-3},
  year = {2019},
  month = oct,
  publisher = {Springer Science and Business Media {LLC}},
  volume = {10},
  number = {1},
  author = {Yao Shen and Changle Liu and Yayuan Qin and Shoudong Shen and Yao-Dong Li and Robert Bewley and Astrid Schneidewind and Gang Chen and Jun Zhao},
  title = {Intertwined dipolar and multipolar order in the triangular-lattice magnet {TmMgGaO}4},
  pages = {4530},
  journal = {Nat. Commun.}
}

@article{Hu2020Berezinskii,
  doi = {10.1038/s41467-020-19380-x},
  url = {https://doi.org/10.1038/s41467-020-19380-x},
  year = {2020},
  month = nov,
  publisher = {Springer Science and Business Media {LLC}},
  volume = {11},
  number = {1},
  author = {Ze Hu and Zhen Ma and Yuan-Da Liao and Han Li and Chunsheng Ma and Yi Cui and Yanyan Shangguan and Zhentao Huang and Yang Qi and Wei Li and Zi Yang Meng and Jinsheng Wen and Weiqiang Yu},
  title = {Evidence of the Berezinskii-Kosterlitz-Thouless phase in a frustrated magnet},
  pages = {5631},
  journal = {Nat. Commun.}
}

@article{Dun2021neutron,
  doi = {10.1103/physrevb.103.064424},
  url = {https://doi.org/10.1103/physrevb.103.064424},
  year = {2021},
  month = feb,
  publisher = {American Physical Society ({APS})},
  volume = {103},
  number = {6},
  pages = {064424},
  author = {Zhiling Dun and Marcus Daum and Raju Baral and Henry E. Fischer and Huibo Cao and Yaohua Liu and Matthew B. Stone and Jose A. Rodriguez-Rivera and Eun Sang Choi and Qing Huang and Haidong Zhou and Martin Mourigal and Benjamin A. Frandsen},
  title = {Neutron scattering investigation of proposed Kosterlitz-Thouless transitions in the triangular-lattice Ising antiferromagnet TmMgGaO4},
  journal = {Phys. Rev. B}
}

@article{Momma2011vesta,
author = "Momma, Koichi and Izumi, Fujio",
title = "{{\it VESTA3} for three-dimensional visualization of crystal, volumetric and morphology data}",
journal = "J. Appl. Crystallogr.",
year = "2011",
volume = "44",
number = "6",
pages = "1272--1276",
month = "Dec",
doi = {10.1107/S0021889811038970},
url = {https://doi.org/10.1107/S0021889811038970},
}

@misc{Coelho2020,   
    title = {TOPAS Academic v4.1 technical reference},   
    url = {http://www.topas-academic.net},   
    author = {A. A. Coelho},   
    year = {2020},   
    note = {} 
}

@article{Coelho2018,
  doi = {10.1107/s1600576718000183},
  url = {https://doi.org/10.1107/s1600576718000183},
  year = {2018},
  month = feb,
  publisher = {International Union of Crystallography ({IUCr})},
  volume = {51},
  number = {1},
  pages = {210--218},
  author = {Alan A. Coelho},
  title = {TOPAS and TOPAS-Academic: an optimization program integrating computer algebra and crystallographic objects written in  C$\mathplus$$\mathplus$},
  journal = {J. Appl. Crystallogr.}
}

@article{liu2020intrinsic,
  title={Intrinsic quantum Ising model on a triangular lattice magnet Tm Mg Ga O 4},
  author={Liu, Changle and Huang, Chun-Jiong and Chen, Gang},
  journal={Physical Review Research},
  volume={2},
  number={4},
  pages={043013},
  year={2020},
  publisher={APS}
}

@article{PhysRevB.68.104409,
  title = {Interplay of quantum and thermal fluctuations in a frustrated magnet},
  author = {Isakov, S. V. and Moessner, R.},
  journal = {Phys. Rev. B},
  volume = {68},
  issue = {10},
  pages = {104409},
  numpages = {9},
  year = {2003},
  month = {Sep},
  publisher = {American Physical Society},
  doi = {10.1103/PhysRevB.68.104409},
  url = {https://link.aps.org/doi/10.1103/PhysRevB.68.104409}
}

@article{Molpro,
	author = {Werner, Hans-Joachim and Knowles, Peter J. and Knizia, Gerald and Manby, Frederick R. and Sch{\"u}tz, Martin},
	title = {Molpro: a general-purpose quantum chemistry program package},
	journal = {WIREs Comput. Mol. Sci.},
	volume = {2},
	number = {2},
	pages = {242-253},
	doi = {https://doi.org/10.1002/wcms.82},
	url = {https://wires.onlinelibrary.wiley.com/doi/abs/10.1002/wcms.82},
	year = {2012}
}

@article{Cao_Dolg_1,
author = {Cao,Xiaoyan  and Dolg,Michael },
title = {Valence basis sets for relativistic energy-consistent small-core lanthanide pseudopotentials},
journal = {J. Chem. Phys.},
volume = {115},
number = {16},
pages = {7348-7355},
year = {2001},
doi = {10.1063/1.1406535},
URL = {https://doi.org/10.1063/1.1406535}
}

@article{Dolg1989,
        author={Dolg, M. and Stoll, H. and Savin, A. and Preuss, H.},
        title={Energy-adjusted pseudopotentials for the rare earth elements},
        journal={Theor. Chim. Acta},
        year={1989},
        month={May},
        day={01},
        volume={75},
        number={3},
        pages={173-194},
        issn={1432-2234},
        doi={10.1007/BF00528565}
}

@Article{Dolg1993,
author={Dolg, M.
and Stoll, H.
and Preuss, H.},
title={A combination of quasirelativistic pseudopotential and ligand field calculations for lanthanoid compounds},
journal={Theor. Chim. Acta},
year={1993},
month={Jun},
day={01},
volume={85},
number={6},
pages={441-450},
issn={1432-2234},
doi={10.1007/BF01112983}
}

@article{expt_KTm,
  title = {Exchange-renormalized crystal field excitations in the quantum {I}sing magnet {KTmSe}$_2$},
  author = {Zheng, Shiyi and Wo, Hongliang and Gu, Yiqing and Luo, Rui Leonard and Gu, Yimeng and Zhu, Yinghao and Steffens, Paul and Boehm, Martin and Wang, Qisi and Chen, Gang and Zhao, Jun},
  journal = {Phys. Rev. B},
  volume = {108},
  issue = {5},
  pages = {054435},
  numpages = {9},
  year = {2023},
  month = {Aug},
  publisher = {American Physical Society},
  doi = {10.1103/PhysRevB.108.054435},
  url = {https://link.aps.org/doi/10.1103/PhysRevB.108.054435}
}

@misc{artifact,
    key = {},
    note ={The artifact originates from a faulty calibration file of the platform thermometer resistance provided by the manufacturer. The problem was fixed before proceeding with the erbium and lutetium based compounds.}
}

@article{NaTmSe2,
  title = {Interplay between crystal field and magnetic anisotropy in the triangular-lattice antiferromagnet ${\mathrm{NaTmTe}}_{2}$},
  author = {Zheng, Shiyi and Gu, Yiqing and Gu, Yimeng and Kao, Zeyu and Wang, Qisi and Wo, Hongliang and Zhu, Yinghao and Liu, Feiyang and Wu, Liusuo and Sheng, Jieming and Chang, Johan and Ohira-Kawamura, Seiko and Murai, Naoki and Niedermayer, Christof and Mazzone, Daniel Gabriel and Chen, Gang and Zhao, Jun},
  journal = {Phys. Rev. B},
  volume = {109},
  issue = {7},
  pages = {075159},
  numpages = {8},
  year = {2024},
  month = {Feb},
  publisher = {American Physical Society},
  doi = {10.1103/PhysRevB.109.075159},
  url = {https://link.aps.org/doi/10.1103/PhysRevB.109.075159}
}

@article{sengupta2003specific,
  title={Specific heat of quasi-two-dimensional antiferromagnetic Heisenberg models with varying interplanar couplings},
  author={Sengupta, Pinaki and Sandvik, Anders W and Singh, Rajiv RP},
  journal={Physical Review B},
  volume={68},
}

@article{liu2018selective,
  title={Selective measurements of intertwined multipolar orders: Non-Kramers doublets on a triangular lattice},
  author={Liu, Changle and Li, Yao-Dong and Chen, Gang},
  journal={Physical Review B},
  volume={98},
  number={4},
  pages={045119},
  year={2018},
  publisher={APS}
}

@article{PhysRevB.103.104416,
  title = {Phase diagram of the quantum Ising model on a triangular lattice under external field},
  author = {Da Liao, Yuan and Li, Han and Yan, Zheng and Wei, Hao-Tian and Li, Wei and Qi, Yang and Meng, Zi Yang},
  journal = {Phys. Rev. B},
  volume = {103},
  issue = {10},
  pages = {104416},
  numpages = {10},
  year = {2021},
  month = {Mar},
  publisher = {American Physical Society},
  doi = {10.1103/PhysRevB.103.104416},
  url = {https://link.aps.org/doi/10.1103/PhysRevB.103.104416}
}

@article{Klintenberg_et_al,
	title = {Accurate crystal fields for embedded cluster calculations},
	journal = {Comp. Phys. Commun.},
	volume = {131},
	number = {1},
	pages = {120-128},
	year = {2000},
	issn = {0010-4655},
	doi = {https://doi.org/10.1016/S0010-4655(00)00071-0},
	url = {https://www.sciencedirect.com/science/article/pii/S0010465500000710},
	author = {M. Klintenberg and S.E. Derenzo and M.J. Weber}
}

@book{olsen_bible,
	address = {Chichester},
	author = {Helgaker, T and J{\o}rgensen, P and Olsen, J},
	publisher = {John Wiley \& Sons},
	title = {Molecular Electronic Structure Theory},
	year = 2000
}

@article{MRCI_Molpro,
	author={Knowles, Peter J. and Werner, Hans-Joachim},
	title={Internally contracted multiconfiguration-reference configuration interaction calculations for excited states},
	journal={Theor. Chim. Acta},
	year={1992},
	month={Oct},
	day={01},
	volume={84},
	number={1},
	pages={95-103},
	issn={1432-2234},
	doi={10.1007/BF01117405},
	url={https://doi.org/10.1007/BF01117405}
}

@article{Cao_Dolg_2,
        title = {Segmented contraction scheme for small-core lanthanide pseudopotential basis sets},
        journal = {J. Mol. Struc.: THEOCHEM},
        volume = {581},
        number = {1},
        pages = {139-147},
        year = {2002},
        issn = {0166-1280},
        doi = {https://doi.org/10.1016/S0166-1280(01)00751-5},
        url = {https://www.sciencedirect.com/science/article/pii/S0166128001007515},
        author = {Xiaoyan Cao and Michael Dolg}
}

@article{Stoll_et_al,
author = {Stoll, Hermann and Metz, Bernhard and Dolg, Michael},
title = {Relativistic energy-consistent pseudopotentials$-${R}ecent developments},
journal = {J. Comput. Chem.},
volume = {23},
number = {8},
pages = {767-778},
doi = {https://doi.org/10.1002/jcc.10037},
year = {2002}
}

@article{ORCA_5,
author = {Neese, Frank},
title = {Software update: The {ORCA} program system$-${V}ersion 5.0},
journal = {WIREs Comput. Mol. Sci.},
volume = {12},
number = {5},
pages = {e1606},
doi = {https://doi.org/10.1002/wcms.1606},
year = {2022}
}

@article{ORCA_5_1,
author = {Neese,Frank  and Wennmohs,Frank  and Becker,Ute  and Riplinger,Christoph },
title = {The {ORCA} quantum chemistry program package},
journal = {J. Chem. Phys.},
volume = {152},
number = {22},
pages = {224108},
year = {2020},
doi = {10.1063/5.0004608}
}

@article{Derenzo_et_al,
	author = {Derenzo,Stephen E.  and Klintenberg,Mattias K.  and Weber,Marvin J. },
	title = {Determining point charge arrays that produce accurate ionic crystal fields for atomic cluster calculations},
	journal = {J. Chem. Phys.},
	volume = {112},
	number = {5},
	pages = {2074-2081},
	year = {2000},
	doi = {10.1063/1.480776},
	URL = {https://doi.org/10.1063/1.480776}
}

@article{NEVPT2,
author = {Angeli,C.  and Cimiraglia,R.  and Evangelisti,S.  and Leininger,T.  and Malrieu,J.-P. },
title = {Introduction of n-electron valence states for multireference perturbation theory},
journal = {J. Chem. Phys.},
volume = {114},
number = {23},
pages = {10252-10264},
year = {2001},
doi = {10.1063/1.1361246}
}

@article{Kuechle1991,
author = { W.   K\"uchle  and  M.   Dolg  and  H.   Stoll  and  H.   Preuss },
title = {Ab initio pseudopotentials for Hg through Rn},
journal = {Mol. Phys.},
volume = {74},
number = {6},
pages = {1245-1263},
year  = {1991},
doi = {10.1080/00268979100102941}
}

@article{NaCeO2,
  title = {Crystal-field effects competing with spin-orbit interactions in {NaCeO}$_{2}$},
  author = {Bhattacharyya, P. and R\"o\ss{}ler, U. K. and Hozoi, L.},
  journal = {Phys. Rev. B},
  volume = {105},
  issue = {11},
  pages = {115136},
  numpages = {4},
  year = {2022},
  month = {Mar},
  publisher = {American Physical Society},
  doi = {10.1103/PhysRevB.105.115136},
  url = {https://link.aps.org/doi/10.1103/PhysRevB.105.115136}
}

@article{LiNaTmS2,
    author = {Boswell, Matt and Peng, Cheng and Xie, Weiwei},
    title = {Magnetic properties of Tm3+ in layered triangular lattices},
    journal = {Journal of Applied Physics},
    volume = {137},
    number = {18},
    pages = {183901},
    year = {2025},
    month = {05},
    issn = {0021-8979},
    doi = {10.1063/5.0268305},
    url = {https://doi.org/10.1063/5.0268305}
}

@article{Co_c.kim,
    author = {Kim, Chaebin and Kim, Sujin and Park, Pyeongjae and Kim, Taehun and Jeong, Jaehong and Ohira-Kawamura, Seiko and Murai, Naoki and Nakajima, Kenji and Chernyshev, A. L. and Mourigal, Martin and Kim, Sung-Jin and Park, Je-Geun},
    title = {Bond-dependent anisotropy and magnon decay in cobalt-based Kitaev triangular antiferromagnet},
    journal = {Nature Physics},
    volume = {19},
    pages = {1624–1629},
    year = {2022},
    month = {11},
    doi = {10.1038/s41567-023-02180-7},
}

@article{Ru_Pritam,
    author = {Bhattacharyya, Pritam and Bogdanov, Nikolay A. and Nishimoto, Satoshi and Wilson, Stephen D. and Hozoi, Liviu},
    title = {NaRuO2: Kitaev-Heisenberg exchange in triangular-lattice setting},
    journal = {npj Quantum Materials},
    volume = {8},
    number = {52},
    year = {2023},
    month = {04},
    doi = {10.1038/s41535-023-00582-7},
}

@article{Ce_Xie,
  title = {Quantum Spin Dynamics Due to Strong Kitaev Interactions in the Triangular-Lattice Antiferromagnet ${\mathrm{CsCeSe}}_{2}$},
  author = {Xie, Tao and Gozel, S. and Xing, Jie and Zhao, N. and Avdoshenko, S. M. and Wu, L. and Sefat, Athena S. and Chernyshev, A. L. and L\"auchli, A. M. and Podlesnyak, A. and Nikitin, S. E.},
  journal = {Phys. Rev. Lett.},
  volume = {133},
  issue = {9},
  pages = {096703},
  numpages = {8},
  year = {2024},
  month = {Aug},
  publisher = {American Physical Society},
  doi = {10.1103/PhysRevLett.133.096703},
  url = {https://link.aps.org/doi/10.1103/PhysRevLett.133.096703}
}

@article{Ce_ortiz,
  title = {Electronic and structural properties of $\mathrm{RbCe}{X}_{2}$ $({X}_{2}:$ ${\mathrm{O}}_{2},$ ${\mathrm{S}}_{2}, \mathrm{SeS}, {\mathrm{Se}}_{2}, \mathrm{TeSe}, {\mathrm{Te}}_{2})$},
  author = {Ortiz, Brenden R. and Bordelon, Mitchell M. and Bhattacharyya, Pritam and Pokharel, Ganesh and Sarte, Paul M. and Posthuma, Lorenzo and Petersen, Thorben and Eldeeb, Mohamed S. and Granroth, Garrett E. and Dela Cruz, Clarina R. and Calder, Stuart and Abernathy, Douglas L. and Hozoi, Liviu and Wilson, Stephen D.},
  journal = {Phys. Rev. Mater.},
  volume = {6},
  issue = {8},
  pages = {084402},
  numpages = {12},
  year = {2022},
  month = {Aug},
  publisher = {American Physical Society},
  doi = {10.1103/PhysRevMaterials.6.084402},
  url = {https://link.aps.org/doi/10.1103/PhysRevMaterials.6.084402}
}

@article{lkqb-3fc7,
  title = {${\mathrm{TlYbSe}}_{2}$ as a member of the $J$=$\frac{1}{2}$ triangular-lattice Yb delafossite family: From spin liquid to field-induced magnetic order},
  author = {Fujii, T. and Pillaca, M. and B\"artl, F. and Sichelschmidt, J. and Luther, S. and Rosner, H. and Strydom, A. M. and Wosnitza, J. and K\"uhne, H. and Doert, Th. and Baenitz, M.},
  journal = {Phys. Rev. B},
  volume = {112},
  issue = {2},
  pages = {024426},
  numpages = {14},
  year = {2025},
  month = {Jul},
  publisher = {American Physical Society},
  doi = {10.1103/lkqb-3fc7},
  url = {https://link.aps.org/doi/10.1103/lkqb-3fc7}
}
\newpage
\appendix
\section*{Appendix}

We employed the SARC2-DKH-QZVP basis set (BS) for the central Er$^{3+}$ 4$f^{11}$ ion \cite{Aravena_et_al} and all-electron relativistic BSs of triple-$\zeta$ quality for the Se ligands \cite{Dixon_Se} of the ErSe$_6$ octahedron; within the buffer cationic region between the reference ErSe$_6$ octahedron and the point-charge matrix, large-core pseudopotentials were utilized for the six Er$^{3+}$ \cite{Dolg_Stoll_Preuss} and 12 Tl$^{+}$ \cite{Kuechle1991} species.

Negligible changes in the relative energies were observed if only 30 spin quartets and 43 spin doublets instead of all possible states associated with the Er$^{3+}$ 4$f^{11}$ configuration were considered in the quantum chemical calculations (see Table \ref{tab:excited_Er_comp}).

For the central Tm$^{3+}$ 4$f^{12}$ ion energy-consistent quasirelativistic pseudopotentials (ECP28MWB) \cite{Dolg_Stoll_Preuss} and Gaussian-type valence BSs of effective quadruple-$\zeta$ quality \cite{Cao_Dolg_1,Cao_Dolg_2} were employed, whereas we used relativistic pseudopotentials (ECP46MDF) \cite{Stoll_et_al} along with (6s6p)/[4s4p] valence BSs for the six Se ligands of the central TmSe$_6$ octahedron.

Large-core pseudopotentials including the 4$f$ subshell in the core (ECP58MWB) \cite{Dolg1989,Dolg1993} were adopted for the six adjacent Tm sites, along with (4s3p)/[2s1p] valence BSs. We considered large-core pseudopotentials for the 12 Tl nearest neighbors (ECP78MWB) \cite{Kuechle1991} with (4s3p)/[2s1p] valence BSs.
\newpage

\begin{table}[h]
\caption{
Low-energy 4$f$-site multiplet structure in TlErSe$_2$ ($J\!=\!15/2$ -like levels);
relative energies of the lowest eight SO doublets (meV).
All possible states derived from the 4$f^{11}$ configuration entered the SO treatment.
}
\begin{tabular}{l l r}
\hline
\hline
\\[-0.15cm]
Levels     &CASSCF+SOC         &NEVPT2+SOC         \\
\hline
\\[-0.15cm]
1           &\hspace{1cm}0.0                  &0.0                    \\
\\[-0.20cm]
2           &\hspace{1cm}0.9                  &6.5                    \\
\\[-0.20cm]
3           &\hspace{1cm}5.0                  &7.4                    \\
\\[-0.20cm]
4           &\hspace{1cm}6.0                  &8.7                    \\
\\[-0.20cm]
5           &\hspace{1cm}7.0                  &11.3                   \\
\\[-0.20cm]
6           &\hspace{1cm}18.3                 &30.4                   \\
\\[-0.20cm]
7           &\hspace{1cm}20.8                 &32.4                   \\
\\[-0.20cm]
8           &\hspace{1cm}22.3                 &33.4                  \\
\hline
\hline
\end{tabular}
\label{tab:excited_Er}
\end{table}
\begin{table}[h]
\caption{
Low-energy 4$f$-site multiplet structure in TlTmSe$_2$ ($J\!=\!6$ -like levels);
relative energies in meV. 
The lowest 11 spin triplets and nine spin singlets
derived from the 4$f^{12}$ configuration entered the SO treatment.
The levels joined by brackets are doublets not split by the crystal field.
}
\begin{tabular}{l l r}
\hline
\hline
\\[-0.15cm]
Levels       \       &CASSCF+SOC   \          &MRCI+SOC  \\
\hline
\\[-0.15cm]
1                    &\hspace{1cm}0.0             &0.0    \\
\\[-0.20cm]
2                    &\hspace{1cm}1.2             &0.8     \\
\\[-0.20cm]
3 $\rceil$           &\hspace{1cm}1.9             &3.5     \\
4 $\rfloor$ \\
\\[-0.20cm]
5 $\rceil$           &\hspace{1cm}3.3             &5.1 \\
6 $\rfloor$ \\
\\[-0.20cm]
7                    &\hspace{1cm}6.8             &9.0 \\
\\[-0.20cm]
8                    &\hspace{1cm}10.2            &12.9 \\
\\[-0.20cm]
9                    &\hspace{1cm}14.0            &18.7 \\
\\[-0.20cm]
10$\rceil$           &\hspace{1cm}14.7            &19.1 \\
11$\rfloor$ \\
\\[-0.20cm]
12$\rceil$           &\hspace{1cm}17.2            &21.7 \\
13$\rfloor$ \\
\\[-0.20cm]
\hline
\hline
\end{tabular}
\label{tab:excited_TlTm}
\end{table}
\begin{table}[h]
\caption{
Low-energy 4$f$-site multiplet structure (meV) in TlErSe$_2$ at NEVPT2+SOC level,
dependence on the number of states considered in the SO computations.
Q and D stand for quartet and doublet spin multiplicities,
respectively.
}
\begin{tabular}{l c r}
\hline
\hline
\\[-0.15cm]
Levels       \        &(30 Q + 43 D)\footnote{Orbitals as employed in the adjacent column were used.}
                                           &(35 Q + 112 D) \\
\hline
\\[-0.15cm]
1                     &0.0                 &0.0            \\
\\[-0.20cm]
2                     &6.5                 &6.5          \\
\\[-0.20cm]
3                     &7.4                 &7.4          \\
\\[-0.20cm]
4                     &8.6                 &8.7          \\
\\[-0.20cm]
5                     &11.2                &11.3         \\
\\[-0.20cm]
6                     &30.4                &30.4         \\
\\[-0.20cm]
7                     &32.4                &32.4         \\
\\[-0.20cm]
8                     &33.4                &33.4         \\
\hline
\hline
\end{tabular}
\label{tab:excited_Er_comp}
\end{table}
\begin{table}[h]
\caption{
Low-energy 4$f$-site multiplet structure (meV) in TlTmSe$_2$ at CASSCF+SOC level, dependence
on the number of states considered in the SO computations (relative energies in meV).
T and S stand for triplet and singlet spin multiplicities,
respectively.
}
\begin{tabular}{l c r}
\hline
\hline
\\[-0.15cm]
Levels      \        &(11 T + 9 S)\footnote{Orbitals as employed in the adjacent column were used.}
                                          \   &(21 T + 28 S)       \\
\hline
\\[-0.15cm]
1                    &0.0             &0.0    \\
\\[-0.20cm]
2                    &1.2             &1.2     \\
\\[-0.20cm]
3 $\rceil$           &1.9             &1.9     \\
4 $\rfloor$ \\
\\[-0.20cm]
5 $\rceil$           &3.3             &3.2 \\
6 $\rfloor$ \\
\\[-0.20cm]
7                    &6.8             &6.6 \\
\\[-0.20cm]
8                    &10.2            &10.0 \\
\\[-0.20cm]
9                    &14.0            &13.8 \\
\\[-0.20cm]
10$\rceil$           &14.7            &14.4 \\
11$\rfloor$ \\
\\[-0.20cm]
12$\rceil$           &17.2            &16.9 \\
13$\rfloor$ \\
\\[-0.20cm]
\hline
\hline
\end{tabular}
\label{tab:excited_TlTm_CAS}
\end{table}

\end{document}